\documentclass[12pt]{aa}
\usepackage[english]{babel}
\usepackage{graphicx}

\onecolumn
\newcommand{\ab}{Astrophys. Bull. }
\newcommand{\bsao}{Bull Spec. Astrophys. Observatory}
\newcommand{\arep}{Astron. Rep. }
\newcommand{\alet}{Astron. Let. }
\newcommand{\araa}{Ann. Rev. Astron. Astrophys. }
\newcommand{\mnras}{Mon. Not. R. Astron. Soc. }
\newcommand{\apj}{Astrophys. J. }

\newcommand{\aj}{Astron. J. }
\newcommand{\aaa}{Astron and Astrophys.}
\newcommand{\aas}{Astron and Astrophys. Suppl.}

\newcommand{\pasj}{Publ. Astron. Soc. Japan }

\begin{document}

\title{Unity and Diversity of Yellow Hypergiants Family}
\author{V.G.~Klochkova}
%\email{Valentina.R11@yandex.ru}
\institute{Special Astrophysical Observatory RAS, Nizhnij Arkhyz,  369167 Russia}

\date{\today} 

\abstract{We summarize the results of long-term spectral monitoring of yellow
hypergiants (YHGs) of  northern hemisphere with a $R\ge60\,000$
resolution. The spectra of these F--G stars of extremely high luminosity,
compactly located at the top of the Hertzsprung--Russell diagram
revealed a variety of spectral features: various types of
H$\alpha$ profile, the presence (or absence) of forbidden and
permitted emissions, as well as circumstellar components.
Variability of spectral details of various nature is studied.
Absolute luminosity, circumstellar envelope expansion rate and
amplitude of pulsations are determined. The reliability of the YHG
status for V1427\,Aql is reliably  confirmed; manifestations of a
significant dynamic instability of the upper layers of the
atmosphere of $\rho$\,Cas after the 2017 ejection and
stratification of its gas envelope are registered; the lack of
companion in the system of the V509\,Cas hypergiant is proven; a
conclusion is made that the V1302\,Aql hypergiant is approaching to 
the low-temperature boundary of the Yellow Void.
\keywords{stars: massive--stars: evolution--stars: atmospheres--stars: opical spectra}
}

\titlerunning{\it Unity and diversity of yellow hypergiants family}  
\authorrunning{\it Klochkova}

\maketitle

\section{Introduction}

The upper part of the Hertzsprung--Russell diagram (hereinafter,
HRD), near the   Humphreys--Davidson luminosity
limit [\cite{Hump1979}] is populated by the most
massive stars in the advanced stages of evolution: luminous blue
variables (LBVs), Wolf--Rayet stars, supergiants with the B[e]
phenomenon, hypergiants and other unstable high-luminosity stars
(hereinafter, HLSs) with emissions in the spectra. During their
previous evolution, these stars were loosing mass due to a quiet
outflow and owing to stellar wind, the rate of which in some
phases reaches critical values, entering the mode of ejection of
the surface layers of matter. These stars, with an initial mass on
the Main Sequence in the range of $20$--$60~\mathcal{M}_{\sun}$,
being at an advanced stage of evolution after the stage of red
supergiants and having lost a significant fraction of their mass,
have extended outflowing atmospheres and structured gas and dust
envelopes, the presence of which is manifested primarily in the
complex nature of  spectral energy  distribution (SED), as well as
in the features of optical, IR and radio
spectra~[\cite{Jager1998}].
The main groups of massive evolved stars of different masses 
that will be considered or mentioned in the text: red supergiants (RSGs),
yellow hypergiants (YHGs), LBVs, blue supergiants (BSGs), hot
supergiants with a B[e] phenomenon, Wolf-Rayet stars (WR). The top
part of the HRD clearly presents the population density of stars
of the above types in the stellar population of  M\,31 and M\,33
galaxies~[\cite{Hump2017}]. The limitedness of the
sample of YHGs compared to the number of their immediate
predecessors---RSGs is in a good light here. Probable immediate
descendants of yellow hypergiants are located in the HRD areas,
populated by hot supergiants of various types (BSGs, B[e]SGs,
lower-luminosity LBVs).

%\begin{figure*}[t!]
%\includegraphics[angle=-90,width=0.6\textwidth,bb=-270 -430 890 1200,clip]{Fig1.eps} %HRD_che.ps
%\caption{The main phases of the far-evolved massive stars on the
%HRD. The figure is published with the permission of M.~Kraus,
%Ondrejov Astronomical Institute, Czech Republic.} 
%\label{HRD}
%\end{figure*}

\begin{table*}[]
\label{objects} 
\caption{Selected parameters of the discussed
YHGs: mass loss rate d$\mathcal{M}_{\sun}$/d$t$, the presence of a
circumstellar disk, according to~[\cite{Aret510}] based on the
presence of   [O\,I] and [Ca\,II] emissions; the  distance
corresponding to parallax from~[\cite{Gaia}]}
\begin{tabular}{ l|c|c|c|c }
\hline
 ~~~Star  &IR source &$\log($d$\mathcal{M}_{\sun}/$d$t)$,  & Presence & $d$, kpc \\
          &         & $\mathcal{M}_{\sun}$\,year$^{-1}$    & of disk  &          \\
\hline
V1427\,Aql &RAFGL\,2343 / IRAS\,19114+0002& $-2.5 \div -4.3$~[\cite{Castro}]&+~[\cite{Castro}]& $3.22 \pm0.16$ \\     % pi=0.3102
%           &&  & & \\ % 0.0512
$\rho$\,Cas&RAFGL\,3173 / IRAS\,23518+5713&$-4.85$~[\cite{Jager1998}]   &+~[\cite{Aret510}]& $1.05\pm0.21$\\   %0.9470
%           &&&& \\  %  0.2021
V509\,Cas  &IRC+60379 / IRAS\,22579+5640  & $-4.92$~[\cite{Jager1998}]  &+~[\cite{Aret510}]& $4.81\pm0.43$\\   %0.2078
%           &&    & &   \\    %0.0899
V1302\,Aql &IRC+10420 / IRAS\,19244+1115  &$-3.30$~[\cite{Jager1998}] &+~[\cite{Jones93}]& $1.72\pm0.28$\\   %0.5824
%           &&&& \\  %0.1620
\hline
\end{tabular}
\end{table*}

This paper focuses on the group of dynamically unstable F--G stars
of extremely high luminosity with specific spectral features.
Instability manifests itself in the pulsations, stellar wind with
a high  mass loss rate (typical values of
d$\mathcal{M}_{\sun}/$d$t$, see Table~\ref{objects}), as well as
in the recurrent episodes of ejection of large masses of matter,
which leads to a temporary decrease in stellar brightness and a
change in its spectral class. A combination of these features
allowed to distinguish these stars in a small family of yellow
hypergiants. The term `yellow hypergiants' was coined by
de~Jager~[\cite{Jager1998}] instead of a less convenient term
\emph{supersupergiants}. These objects are sometimes referred to
as \emph{cool  hypergiants}~[\cite{Shuster2006}]. According to the
modern ideas about the evolution of massive stars, this stage is
populated by stars with an initial mass in the range of about
$20$--$40~\mathcal{M}_{\sun}$~[\cite{Jager1998,Meynet2007}]. After
the burnout of hydrogen in the core, a massive star spends most of
its further existence  in the stage of supergiants, whose
luminosity is provided by the burning of helium in the core.
Interest in these objects is due, in particular, to the fact that
the observed features, which are mainly registered  owing to the
long-term monitoring of these rapidly evolving objects, serve as a
data source   for testing the models of massive star evolution and
chemical composition of the Galaxy. The understanding of these
final phases of evolution of massive stars is also important in
the issue of supernova precursors (see the paper by Jura et
al.~[\cite{Jura2001}]).

%%%

Currently, less than a dozen yellow hypergiants are identified in
the Galaxy, all of which are listed in the
survey~[\cite{Jager1998}]. A small number of YHGs is
due to the low duration of this evolutionary phase:
de~Jager~[\cite{Jager1998}] indicated the
characteristic duration below $10^5$~years, later
Massey~[\cite{Massey2013}] presumed  an extremely low
characteristic duration, measured only in thousands of years, i.e.
about 0.01\% of a massive star's lifetime. On the HRD, yellow
hypergiants having the luminosity of \mbox{$\log
L/L_{\sun}=5$--$6$} (the luminosity class of Ia$^+$, according to
the MK classification), occupy a limited area near the limiting
luminosity~[\cite{Nieuwen2012}]. In addition to
extremely high luminosities, a specific property of YHGs is their
internal dynamic instability~[\cite{Jager2001}]. An
instability, manifested in the pulsations and in multiple episodes
of giant masses of matter being ejected, provides for the
formation of a powerful and structured circumstellar gas-and-dust
envelope. Variations in the optical thickness of the envelope are
revealed in a zigzag passage of the star's position on the HRD, as
was observed, for example, in the star
V509\,Cas~[\cite{Nieuwen2012}]. Quite expectable for
such massive stars is a detection  in the atmospheres of YHGs of
large nitrogen and sodium
excesses~[\cite{Boyarchuk,IRC1,Sahin}],
the synthesis and dredge-up of which to the near-surface
atmospheric layers occur at the previous stages of evolution of
massive stars.

%%%
In this paper, we compare the observed differences in spectral and
kinematic features, as well as their temporal behavior in four
northern hemisphere yellow hypergiants listed in
Table~\ref{objects}.
Section~\ref{obs} shortly summarizes the methods of
observations and data analysis.
Section~\ref{results} lists the results obtained,
comparing them with those previously published, and in
Section~\ref{conclus} we give our conclusions.

\section{Spectral data}\label{obs}

Long-term spectral monitoring of four YHGs was conducted with the
NES echelle spectrograph~[\cite{nes1,nes2}] placed at the
Nasmyth focus of the 6-m BTA telescope of the Special Astrophysical
Observatory of the Russian Academy of Sciences (SAO RAS).
The NES spectrograph is equipped with a large format
CCD sized 2048$\times$2048 elements (in recent years, a 4608$\times$2048 CCD)
and is fitted   with a three-slice image slicer, which reduces light loss with no   spectral
resolution loss. Each spectral order in the image is reiterated
three times, with a shift along the dispersion of the echelle grating~[\cite{nes2}].
The spectral resolution is $\lambda/\Delta\lambda\ge60\,000$,
the signal-to-noise ratio $S/N>100$ for all the spectra used in this work.
%%%

One-dimensional data were extracted from the 2D echelle spectra
using the especially modified for the  spectrograph's
echelle-frame features ECHELLE context, featured in the ESO MIDAS
reduction system (see the details
in~[\cite{echelle}]). Traces of cosmic particles were
removed by the median averaging of two echelle spectra
successively obtained one after the other. Wavelength calibration
was carried out using the Th\mbox{--}Ar hollow-cathode lamp.
Further reduction including the photometric and positional
measurements was performed with the latest version of the DECH20t
code~[\cite{dech}]. Note that this program code that
we traditionally use to reduce the spectra allows us to measure
radial velocities for separate line profile features. We only use
here the heliocentric velocities $V_r$, the systematic errors of
which, estimated from sharp interstellar components of Na\,I lines
do not exceed 0.25~km~s$^{-1}$ (from a single line), mean random
errors for shallow absorptions are of about~$0.7$~km\,s$^{-1}$ per
line.

%%%

Spectral monitoring of V1427\,Aql was performed at two telescopes:
the SAO RAS 6-m BTA telescope in combination with the NES spectrograph, and the   McDonald
Observatory 2.7-m reflector  with a coude echelle spectrograph.
As  follows from a comparison in~[\cite{Sahin}], material obtained at two
telescopes is homogeneous in spectral resolution.
A detailed identification of features in the spectrum of V1302\,Aql was published
based on the spectral atlas~[\cite{atlas}], in which
the spectra of V1302\,Aql and  MWC\,314 supergiant with the
B[e] phenomenon are thoroughly compared.
Identification of features in the spectra of V509\,Cas,
V1427\,Aql and $\rho$\,Cas is conducted using the
previously published spectral atlas for the F--G stars~[\cite{atlasFG}]
and the VALD database information.

\section{Main parameters  of yellow hypergiants}\label{results}

The yellow hypergiants discussed in this article are listed in
Table~\ref{objects}, which gives the names of the associated IR sources, the most used for
these stars, as well as some
information needed to compare these stars. The mass loss rate
for V1427\,Aql is borrowed from~[\cite{Castro}],
whose authors estimated the parameter by simulating the emission of CO.
Further, the presence in the object of a circumstellar disk is noted according to~[\cite{Aret510}],
based on the presence of forbidden [Ca\,II] and [O\,I] emissions in the spectrum.
The last column of this table  gives distance for each
star based on the parallax according to the DR2
Gaia data~[\cite{Gaia}].
Extremely low distance values for V1427\,Aql, $\rho$\,Cas and V1302\,Aql
are conspicuous here.
The remoteness of these stars obtained by other methods is significantly
larger: about~6~kpc for V1427\,Aql according to the authors of~[\cite{Jura2001}],
about 2.5~kpc for $\rho$\,Cas~[\cite{Shuster2006}] and up to~5--6~kpc for
V1302\,Aql~[\cite{Jones93,Kastner}]. This contradiction serves as an
additional confirmation of limited parallax accuracy
for stars with extended envelopes~[\cite{Xu2019,V1648Aql}].

%%%%%%m table 2
\begin{table*}[]
\caption{Data on YHGs based on the optical spectroscopy data: the
absolute magnitude $M_V$, pulsation  amplitude $\Delta V_r$,
envelope expansion velocity  $V_{\rm exp}$, microturbulent
velocity $\xi_t$, the presence or absence of emission and its
variability in the H$\alpha$ lines, forbidden emissions of $\rm
[Fe\,II]$ and other metals, emissions of the doublet $\rm
[Ca\,II]$ 7291 and 7324~\AA, highly excited $\rm [N\,II]$ emissions} 
\medskip
\begin{tabular}{l|c|c|c|c}
\hline
\multicolumn{1}{r|}{Object}   &  V1427\,Aql   &$\rho$\,Cas   & V509\,Cas    & V1302\,Aql \\
\multicolumn{1}{l|}{Parameter} & [\cite{Sahin}]  &[\cite{rhocas2}]&[\cite{V509Cas}]&[\cite{IRC3}] \\
\cline{2-5} \hline
$W_{\lambda}$(7773), \AA{}  & $2.70$          & $1.86$          & $2.35$          & $2.86$ \\
$M_V$                   &$-8\fm9$ &$-8\fm0$&$-8\fm8$& $\le -9\fm5$ \\
$\log L/L_{\sun}$    & $5.47$           & $5.11$         & $5.43$          & $5.71$ \\
\hline
$\Delta V_r$,  km\,s$^{-1}$ & $11$       &$\ge 10$      & $9$             & $7$  \\
\hline
$V_{\rm exp}$,  km\,s$^{-1}$          &$\approx 11$   & $13$--$23$    & $33$--$40$    & $\approx 40$ \\
\hline
$\xi_t$,  km\,s$^{-1}$       & $6.6$           & $12$           &              & $12$  \\
                     &  $4.76^1$     &    $11.1^2$  & $4.00^1$; $11^2$ & $7.0^2$ \\
\hline H$\alpha$, emission   & in wings     &  in wings   & +
& +   \\ [-5pt]
                     & var           &  var         &  var         & var \\
%\hline
$\rm [Fe\,II]$, emission& --            &  --          &  +           & +  \\
$\rm [Ca\,II]$, emission&  +            &  var$^3$     &  +       & +  \\
$\rm [N\,II]$, emission &  --           &  --          &  +           & --  \\
\hline
\multicolumn{5}{l}{\small\it Footnotes:  1 -- data adopted from~[\cite{Luck2014}],  2  -- from~[\cite{Takeda1998}],  3   --  $\rm [Ca\,II]$  emission is}\\
\multicolumn{5}{l}{\small\it  present in the spectra of  $\rho$\,Cas  only around  the brightness minimum }  \\
\end{tabular}
\label{parameters}
\end{table*}

Table~\ref{parameters} lists the parameters for
YHGs, obtained from spectral observations mainly on the 6-m BTA
telescope. The first line of this table indicates the equivalent
width $W_{\lambda}$(7773) of the oxygen triplet OI\,7773~\AA, the
well-known luminosity criterion for the evolved F--G stars. What
follows is the absolute magnitude $M_V$ corresponding to the
oxygen triplet intensity using the calibration
from~[\cite{Kovtyukh}]. As expected, luminosity of
all four stars is very high and close to the limit. The use of the
intensity of the oxygen triplet  OI\,7773~\AA{} to determine the
luminosity of YHGs in the presence of the Gaia data is due to the
specificity of stars we are investigating. Gaia's high-precision
parallaxes in Table~\ref{objects} may provide exact
distances, but for stars with powerful dust envelopes they do not
provide the necessary accuracy in luminosity estimates. It is
useful to refer here to the study of Xu et
al.~[\cite{Xu2019}], in which a comparison of Gaia
data and VLBI radio spectroscopy data led the authors to conclude
on a smaller  accuracy in the  Gaia data for the stars with dust
envelopes. The luminosity accuracy is also limited by
unreliability of the reddening estimations    for stars with
envelopes. A good illustration of  the difficulty in the reddening
estimation is presented in its detailed study for V509\,Cas
in~[\cite{Nieuwen2012}].

\subsection{Instability of yellow hypergiants}

Hypergiant instability manifests itself in a weak brightness
variability (with an amplitude of about~$0\fm2$--$0\fm5$), which
is usually referred to as a pulsation
type~[\cite{Jager2001}], as well as in the
variability of the profiles of   spectral features that  form  at
different depths of the atmosphere. In the
survey~[\cite{Jager1998}] de~Jager emphasized that
sometimes a star of a very high luminosity is referred to
hypergiants, while a fundamental feature that distinguishes the
spectrum of a supergiant of the Ia$^+$ luminosity class from the
spectrum of an Ia supergiant lies in  large-scale motions in the
atmosphere of the star, manifesting themselves in record-large
absorption widths, as well as the presence of an extended envelope
due to the high mass loss rate.
Table~\ref{parameters} shows the parameters of
YHGs, characterizing the dynamic instability of their atmospheres:
pulsation amplitude $\Delta V_r$, expansion velocity of the gas
envelope $V_{\rm exp}$, as well as the microturbulent velocity
values $\xi_t$, obtained from the homogeneous spectral data in the
studies cited in the table. To make the picture complete, we
attracted the $\xi_t$ values from~[\cite{Takeda1998,Luck2014}]. As
follows from Table~\ref{parameters}, microturbulent
velocity is high for all hypergiants. In case of a hypergiant
V1302\,Aql, for which accretion of matter is detected,
Table~\ref{parameters} lists its velocity. Note the
strikingly close pulsation amplitude values, \mbox{$\Delta
V_r\approx10$~km\,s$^{-1}$},  estimated from the absorptions of
weak and moderate intensity, forming in the deep layers of
atmosphere. It is appropriate to recall here another problem in
the analysis of YHGs spectra, namely, the probable variability of
microturbulent velocity at different depths in their atmospheres.
This effect was first discovered by Boyarchuk et al.
(see~[\cite{Boyarchuk}] and references therein) for
$\rho$\,Cas. However, subsequent studies do not confirm this
result~[\cite{Lobel1992}].

%%%

Being so distinguished by the photometric and spectral properties,
hypergiants belong to stars for which fixation of the evolutionary
stage is difficult. Long-term studies were needed for each of them
before any certainty in this task was to be achieved. For example,
V1302\,Aql and V1427\,Aql had long been attributed to the stars in
the stage after the asymptotic giant branch, inhabited by
intermediate-mass stars with initial masses in the range of
$2$--$9~\mathcal{M}_{\sun}$ at an advanced stage of evolution with
an energetically inert core and two inlayer  sources of nuclear
burning. After a long evolutionary path, these stars are observed
in the final phase of transition to planetary nebulae and
subsequently to white dwarfs, which allows them to be called
protoplanetary nebulae (PPN). However, the data obtained for
V1302\,Aql and V1427\,Aql in recent decades using different
methods[\cite{Hump1979,Hump1997,IRC1,IRC2,Oudm2009,Sahin,IRC3}],
allowed to come to an unambiguous interpretation of the status of
these two stars as YHGs. Note that   the SIMBAD database still
lists for both of these stars  an indication of their belonging to
post-AGB stars. However, two other stars: V509\,Cas and
$\rho$\,Cas are catalogued as long period variables.

%%%

The reverse cases of incorrect classification of HLSs with IR
excess are also known. The situation with the star V1610\,Cyg,
identified with the Egg nebula ($=$\,RAFGL\,2688) can serve as an
example to it. Owing to its high luminosity and the presence of
excess infrared radiation this star was attributed to massive
stars, what was noted in the
survey~[\cite{Jager1998}]. The protoplanetary nebula
phase for Egg has been confirmed by the spectral features in the
radio, infrared and optical spectral ranges indicating the carbon
enrichment of the stellar envelope~[\cite{Bakker97,Loup}].
Later, reduced metallicity in the atmosphere of this star and
anomalies in the abundance of chemical elements, typical for the
PPN stage were detected~[\cite{Egg1}].

%%%
Below, considering the individual characteristics of the spectra of YHGs, we
dwell on the manifestations of the spectral variability of these
objects.

\subsection{Differences in the optical spectra of YHGs}

\begin{figure}[ht!]
%\onelinecaptionstrue
%\captionstyle{normal}
%\setcaptionmargin{5mm}
\includegraphics[angle=0,width=0.8\columnwidth,bb=10 5 545 700,clip]{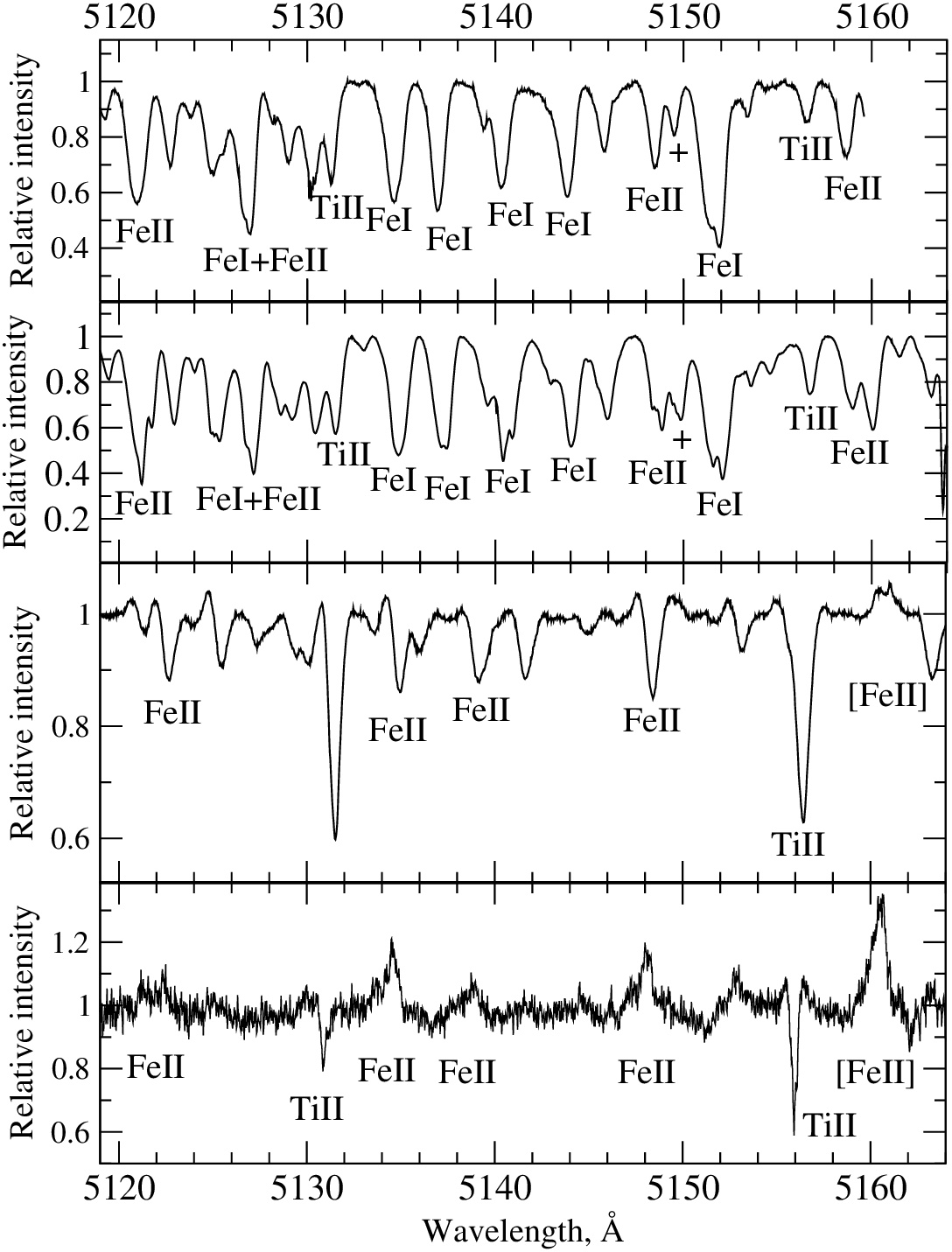} %Frag5150.eps
\caption{A comparison of the $\Delta\lambda=$\,5119--5164~\AA\
interval in the spectra of HD\,179821,
        $\rho$\,Cas, V509\,Cas and V1302\,Aql (from top to bottom,
        in the same order as in Table~\ref{objects}).
        The main absorptions of the fragment are identified at
        the laboratory wavelengths. On the top two panels
        the cross marks the Na\,I\,5149~\AA\ absorption.}
\label{5150}
\end{figure}

\begin{figure}[ht!]
\includegraphics[angle=0,width=0.8\columnwidth,bb=10 25 545 700,clip]{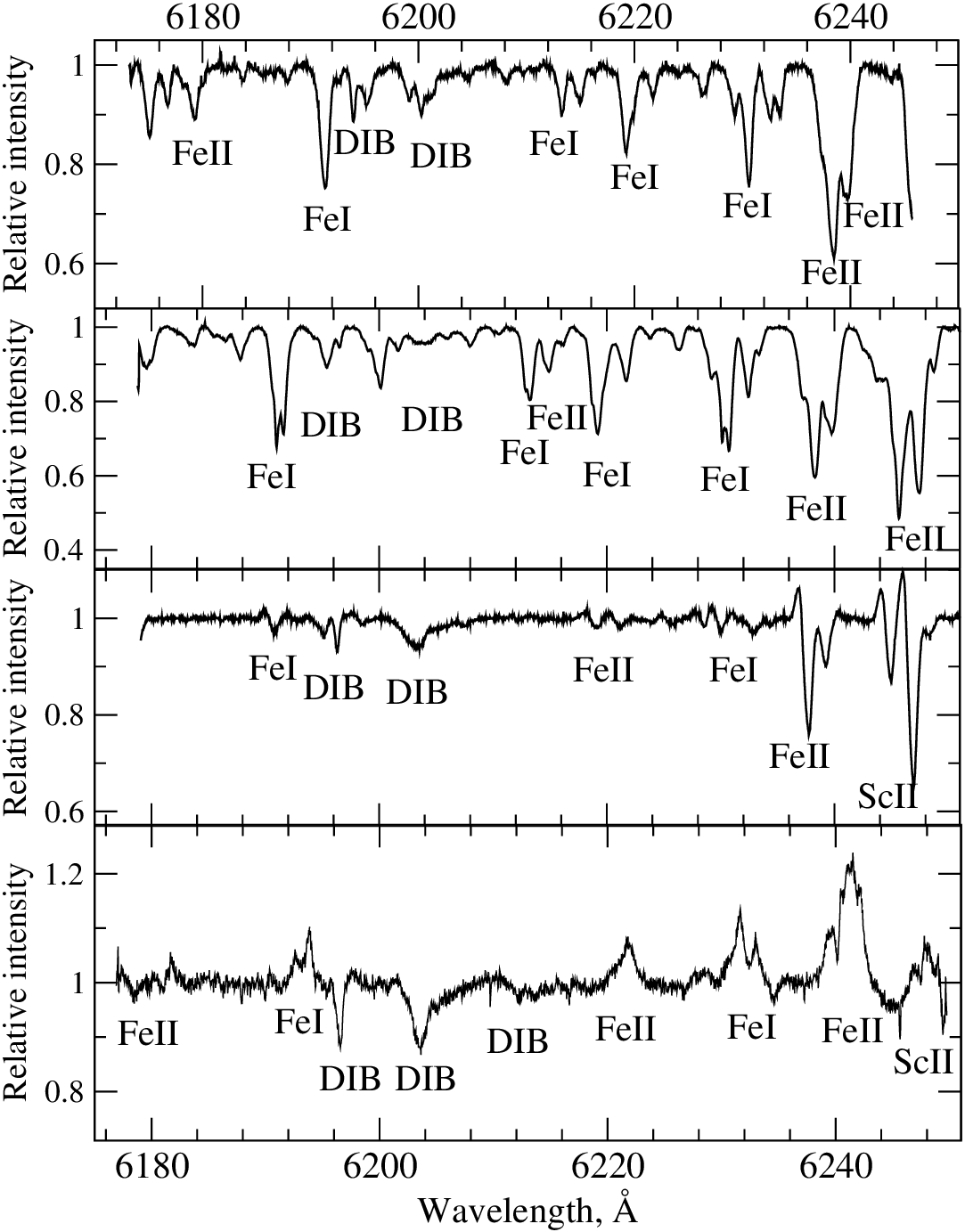} %Frag6200.eps
\caption{The same as in Fig.~\ref{5150}, but for the
\mbox{$\Delta\lambda=$\,6170--6250~\AA} range, containing the
DIBs.} 
\label{6200}
\end{figure}

\begin{figure}[ht!]
\includegraphics[angle=0,width=0.8\columnwidth, bb=20 20 545 750,clip]{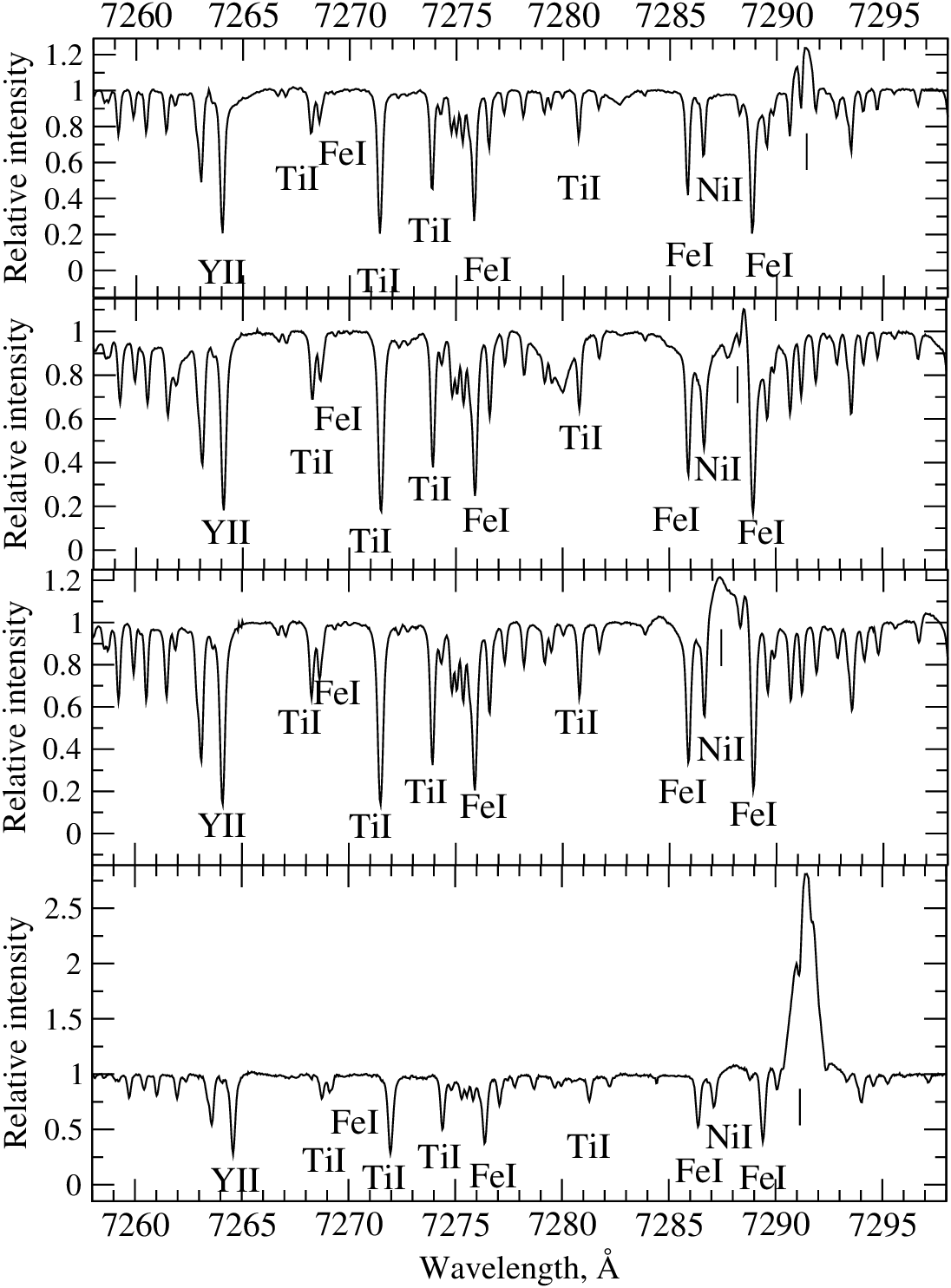}  % Frag7250.eps
\caption{The same as in Fig.~\ref{5150}, but for the
$\Delta\lambda=$\,7258--7298~\AA\ range. For $\rho$\,Cas, a
fragment of the October 1, 2014 spectrum was used, i.e. after the
brightness minimum. Short vertical lines mark the position of the
[Ca\,II]\,7291~\AA\ emission. The spectrum of HD\,179821 is
obtained with a coude echelle spectrograph of the McDonald
Observatory 2.7-m reflector (see paper [\cite{Sahin}] for details.}
\label{7250}
\end{figure}

\begin{figure}[ht!]
\includegraphics[angle=0,width=0.8\columnwidth, bb=20 35 540 695,clip]{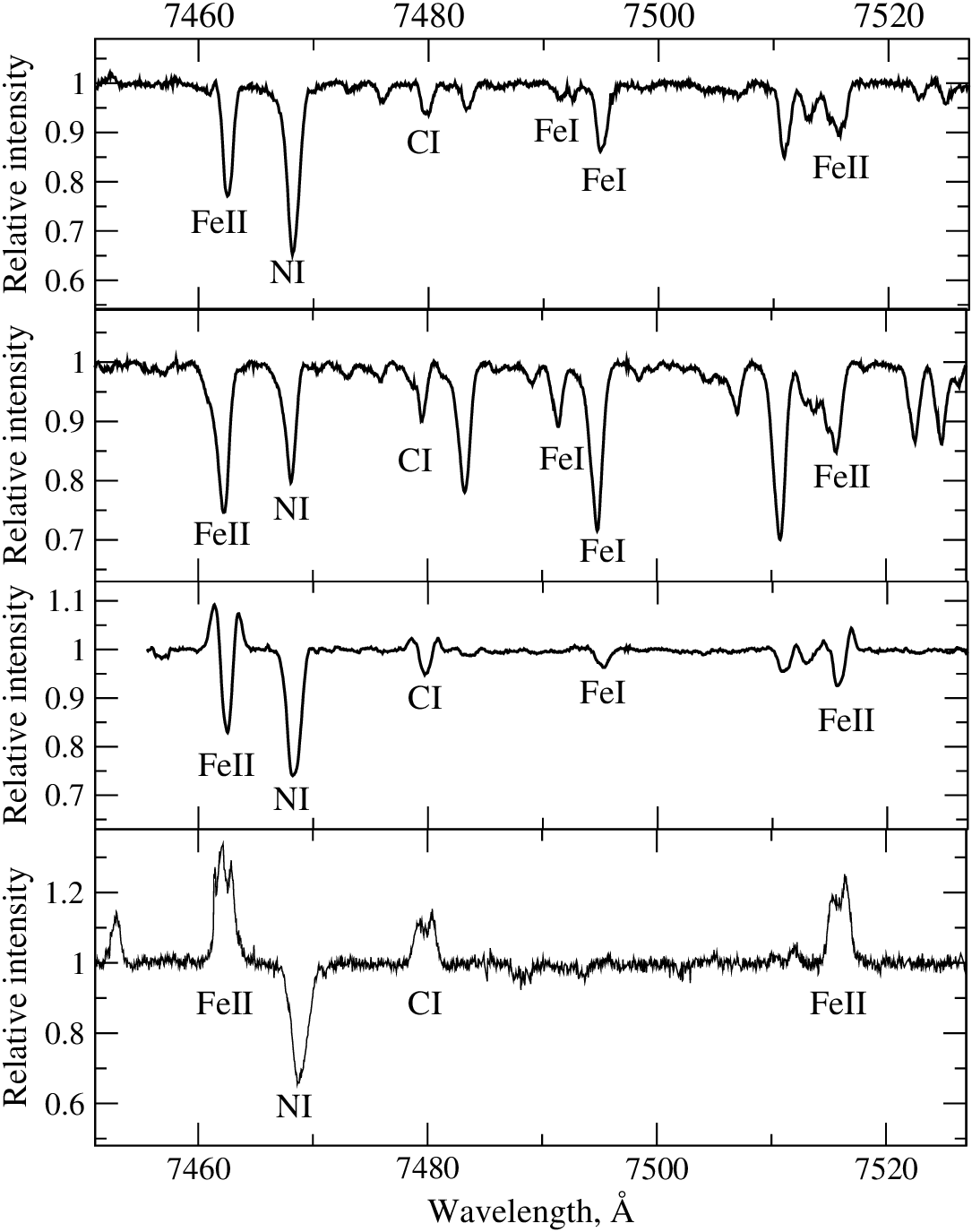} %Frag7450.eps
\caption{The same as in Fig.~\ref{7250}, but for the
\mbox{$\Delta\lambda=$\,7450--7530~\AA} range.} 
\label{7450}
\end{figure}

Given close fundamental parameters of YHGs and in the presence of
common main features in the spectra of representatives of this
family ($\rho$\,Cas, V509\,Cas, V1302\,Aql, HD\,179821), rather
compactly located on the HRD, some observable properties and the
features of their spectra significantly diverse, which is clearly
seen at the fragments presented in
Figs.~\ref{5150},~\ref{6200},~\ref{7250}
and~\ref{7450}. Some differences are also
registered in the diversity of temporal behaviors of the features
of optical spectra, which is largely due to the particularities of
the structure, optical density and kinematics of the envelopes.
Below we consider in more detail the features of optical spectra
of four northern hemisphere YHGs, available for observations with
high spectral resolution and therefore being the most studied.

In addition to the already mentioned large absorption widths, the
principal feature of the optical spectra of YHGs is the presence
of forbidden emissions of atoms and ions formed in the tenuous
outer layers of extended atmospheres and in the circumstellar
medium of these objects.  The [O\,I] and [Ca\,II] emissions, also
available for study from the medium-resolution spectra are studied
well enough  (see~[\cite{Aret2017,Aret510}] and
references in these publications). These emissions are also
recorded  in our spectra as single-peak or two-peak
features~[\cite{atlas, IRC2, IRC3, rhocas2, rhocas3, V509Cas}].

\subsubsection{Hypergiant  V1302\,Aql}

The hypergiant V1302\,Aql is associated with the brightest source
of IR flux IRC+10420. OH-maser radiation is also registered in the
system of this source~[\cite{Giguere}]. Note that
before this the hottest maser sources were associated only with
the class M3 stars. In the framework of the classical
interpretation of a two-peak OH-spectrum, an estimate of the
kinematic distance to the IRC$+$10420 source, $6.8$~kpc was made.
From here followed  an estimate of the absolute magnitude
\mbox{($M_V<-9\fm4$)}, well consistent with our estimate in
Table~\ref{parameters}, unusually high for an F8I
supergiant. To explain the presence of maser sources in the
vicinity of such a hot object a hypothesis of the formation of a
powerful gas and dust envelope at the stage of an M-supergiant
followed by a fast evolution of IRC$+$10420 to the region of
higher temperatures on the HRD~[\cite{Mutel}] was
suggested. The assumption on  the evolution of the star was
confirmed later, based on the observations with high spectral
resolution~[\cite{Oudm1998,IRC2,IRC3}].

\begin{figure}[ht!]
\includegraphics[width=0.3\columnwidth, bb=40 0 350 610,clip]{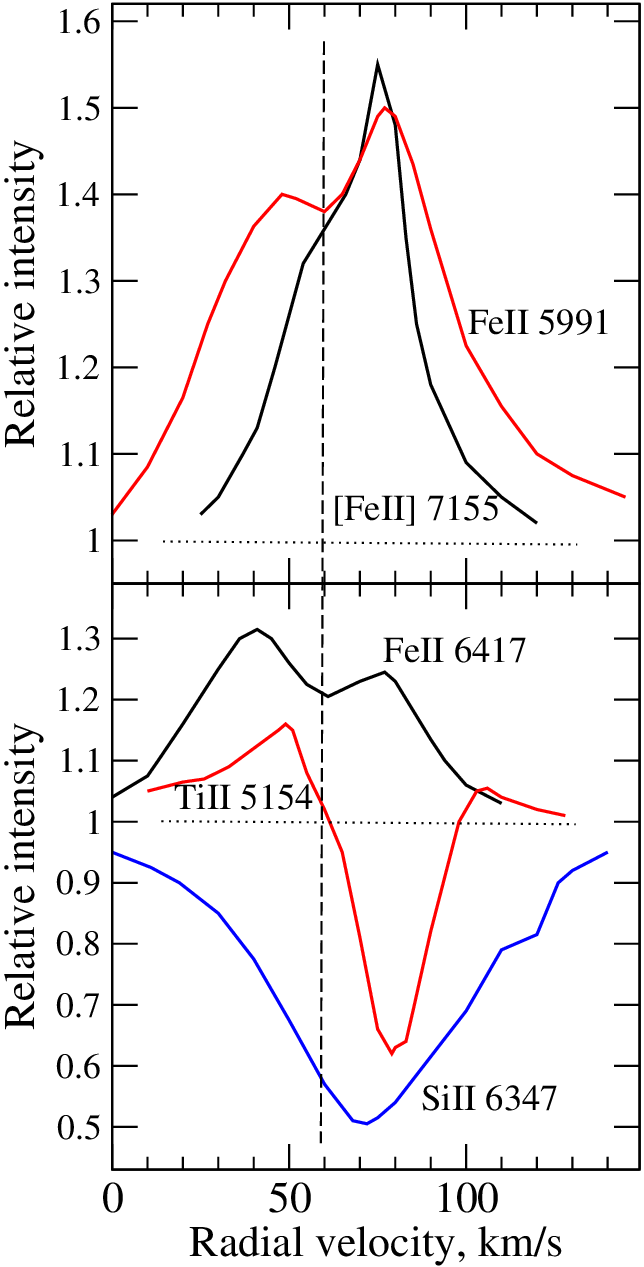}  %IRC_prof.eps
\caption{A variety of line profiles in the spectrum of V1302\,Aql
in 2014. Top panel: a forbidden  [Fe\,II]\,7155~\AA\ emission and
Fe\,II\,5991~\AA\ emission. Bottom panel: Fe\,II\,6417~\AA\
emission, emission--absorption Ti\,II\,5154~\AA\ line  and
Si\,II\,6347~\AA\ absorption. The dotted line describes the
continuum. The vertical dashed line plots the systemic velocity
$V_{\rm sys}\approx60$~km\,s$^{-1}$~[\cite{Oudm1996}].}
\label{IRC3_prof}
\end{figure}

\begin{figure*}[ht!]
\includegraphics[angle=0,width=0.45\textwidth, bb=20 80 545 676,clip]{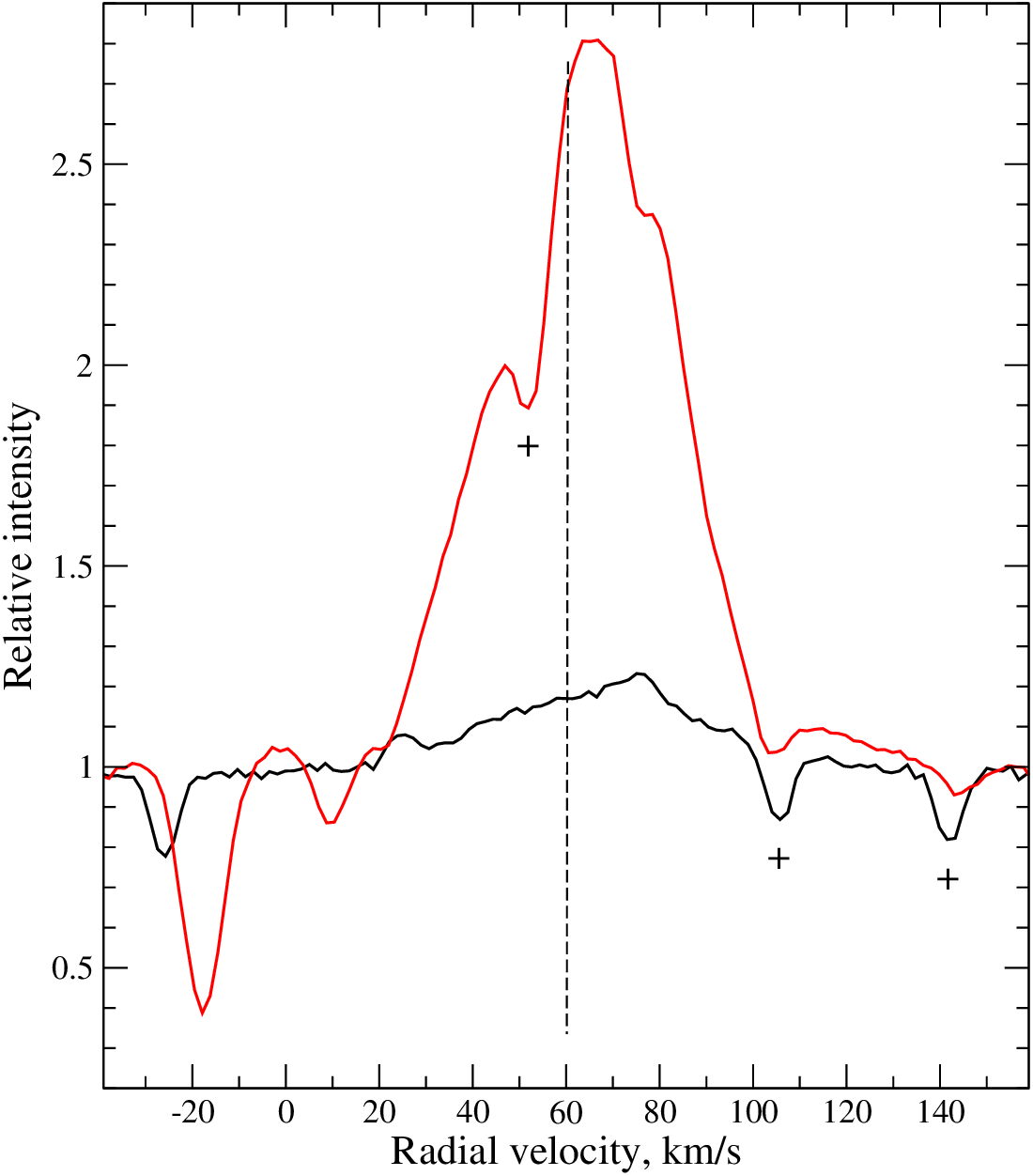}  %IRC_prof_7291_6300.eps
\includegraphics[angle=0,width=0.45\textwidth, bb=20 80 545 676,clip]{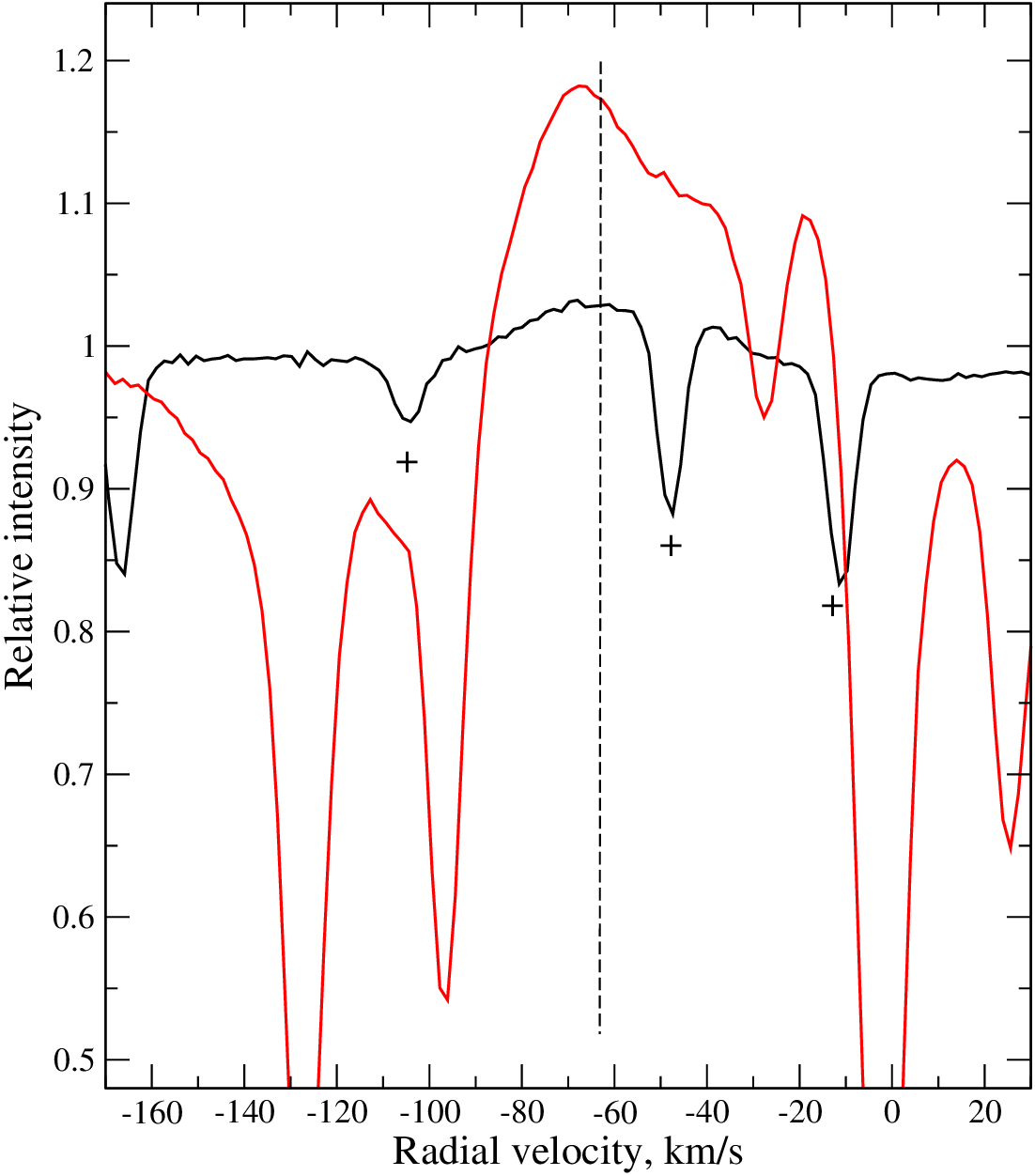}  %V509Cas_prof_7291_6300.eps
\caption{The profiles of  forbidden emissions of
[Ca\,II]\,7291~\AA\ (the red line) and [O\,I]\,6300~\AA\ in the
spectra of V1302\,Aql (left) and  V509\,Cas. Crosses indicate
undeleted telluric features near the emissions. The vertical
dashed line indicates the accepted value of the systemic velocity
$V_{\rm sys}\approx60$~km\,s$^{-1}$ for V1302\,Aql~[\cite{Oudm1996}] 
and  $V_{\rm sys}=-63$~km\,s$^{-1}$  for V509\,Cas~[\cite{V509Cas}].} 
\label{prof_7291_6300}
\end{figure*}

%%%
The optical spectrum of V1302\,Aql is saturated with complex
emission-absorption line profiles,  forbidden emissions of [O\,I],
[Ca\,II], [Fe\,II] and other metals of the iron group, as well as
interstellar bands, DIBs. All their diversity, from pure
absorptions, up to one-peak and/or two-peak forbidden and
permitted emissions, are well illustrated by the spectral
fragments in Figs.~\ref{5150},
\ref{6200}, \ref{7250} and
\ref{7450} and selected line profiles in
Figs.~\ref{IRC3_prof} and
\ref{prof_7291_6300}.

\begin{figure}[ht!]
\includegraphics[angle=0,width=0.5\columnwidth, bb=20 75 545 675,clip]{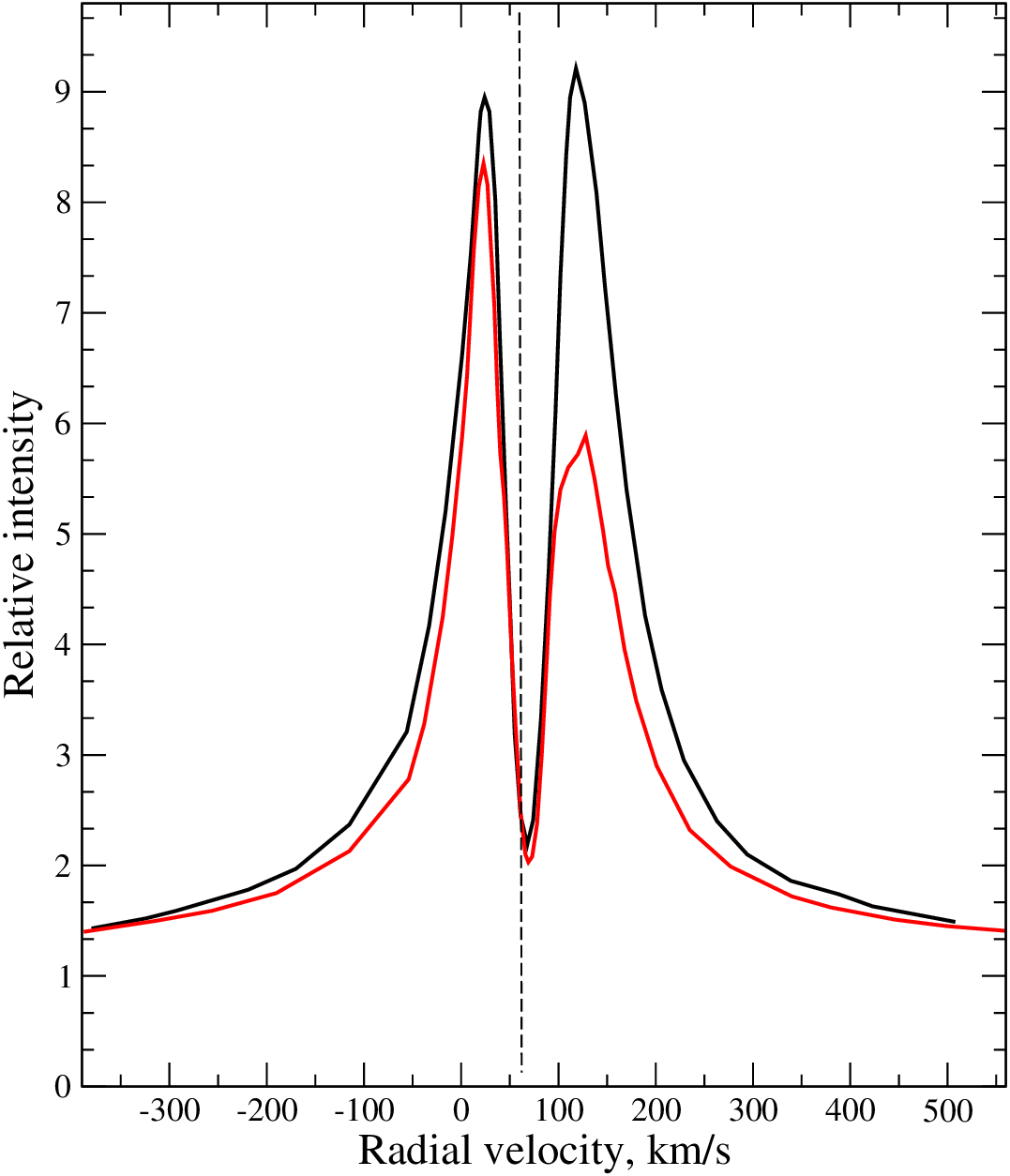}  %IRC_Halpha.eps
\caption{The H$\alpha$ profile in the spectrum of V1302\,Aql in 2007 is
  described by the black line, and in 2014---by a red line. The vertical dashed line  plots 
  the systemic velocity $V_{\rm sys}\approx60$~km\,s$^{-1}$~[\cite{Oudm1996}].}
\label{IRCHalpha}
\end{figure}

Figure~\ref{IRCHalpha} presents  H$\alpha$ profiles
in two  V1302\,Aql spectra, obtained over  different years. In the 2014 spectrum
we can see the  ratio of emission peak intensities,
typical for the entire 20-year observation period at the 6-m
telescope, where the shortwave peak is much more intense than the longwave peak~[\cite{IRC2,Hump2002}].
Only on one date of observations, November 24, 2007, we recorded an unusual shape of the
H$\alpha$ profile, in which the longwave peak   substantially exceeds the shortwave 
one~[\cite{IRC3}]. A comparison of other feature profiles in the spectra of V1302\,Aql 
for the period over 2001--2014 indicates an absence of their significant variability~[\cite{IRC3}].

%%%

One of  decisive arguments confirming the status of a massive
far-evolved HLS for V1302\,Aql, was obtained based on the spectral
data of the 6-m telescope, when the authors
of~[\cite{IRC1}] for the first time determined the
chemical composition of the atmosphere of V1302\,Aql and found a
significant excess of nitrogen. Enrichment of  stellar atmosphere
with nitrogen synthesized by the CNO process deep within the
massive star is a distinctive sign of a large initial mass of the
star. In particular, strong nitrogen absorption, along with the
permitted and forbidden lines of  Fe\,II ions, is contained in one
of the 40 echelle orders of the  V1302\,Aql spectrum, obtained
with the NES spectrograph on  December 8, 2017 and shown at the
bottom panel of Fig.~\ref{7450}. Note  that later
on, the authors of~[\cite{Quintana13}] based on the
observations with the IRAM radio telescope discovered a nitrogen
oxide NO molecule in the circumstellar medium of IRC+10420, which
confirms the nitrogen excess in the atmosphere of its central
star.

%%%

Interest in the hypergiant V1302\,Aql intensified due to the
discovery of a rapid growth of its effective
temperature~[\cite{Oudm1996,IRC1,Oudm1998,IRC2}],
which allowed to assume that the star is rapidly evolving
(probably, to a Wolf--Rayet stage) with a $T_{\rm eff}$ increase
rate of about $120$~K a year. The main conclusion obtained as a
result of long-term spectral monitoring in~[\cite{IRC3}] consists in that this YHG has
entered a phase of a slowdown (or cessation) of effective
temperature growth and approached the border of the Yellow Void on
the HRD. It is obvious that the further monitoring of this star,
observed at the evolutionary transition, whose direction is a
priori difficult to predict, is indispensible. The high spectral
resolution spectroscopy data are necessary to refine the structure
and kinematics of its circumstellar envelope. It seems to us that
the `rain' model proposed by Humphryes et al.~[\cite{Hump1997,Hump2002}] is the
most adequate with regards to the kinematic data.

\subsubsection{$\rho$\,Cas, a yellow hypergiant without a dust envelope}

\begin{figure}[ht!]
\includegraphics[width=0.4\columnwidth, bb=20 70 365 790,clip]{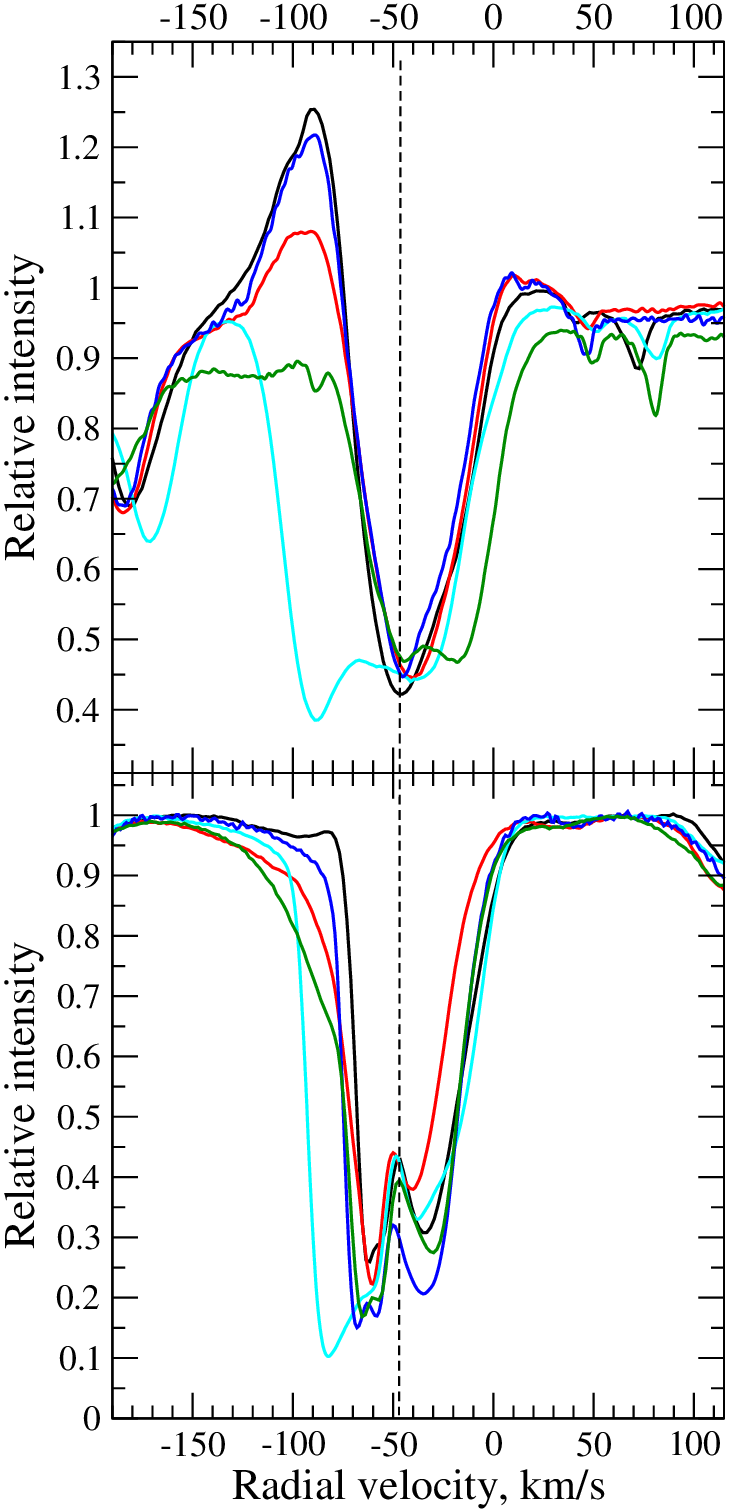}  %RhoCas_profiles_var.eps
\caption{Variable  profiles of H$\alpha$ (top panel) and Ba\,II\,6141~\AA{} in the spectra of 
$\rho$\,Cas, obtained at the BTA during different observational sets: September 30, 2009 (in a
quite state---the black solid line), February 02, 2013 (the
beginning of the 2013 ejection---red), 08/07/2014 (after the 2013
minimum---cyan), February 13, 2017 (before the 2017 ejection---blue), 10/05/2017 
(close to the 2017 minimum---dark green). The vertical dashed line indicates the value of  
$V_{\rm sys}=-47$~km\,s$^{-1}$~[\cite{Lobel1998}].}
\label{RhoCas}
\end{figure}

Peculiarity of the optical spectrum of V1302\,Aql is clearly
manifested in comparison with a related object, a yellow
hypergiant $\rho$\,Cas, whose spectrum, in contrast to the
spectrum of V1302\,Aql, is predominantly absorption, as
illustrated in Figs.~\ref{5150}, \ref{6200}, \ref{7250} and
\ref{7450}. Belonging of $\rho$\,Cas to the group
of YHGs is confirmed by the features of chemical composition of
its atmosphere: the content of CNO elements and a sodium excess
indicate that the star has already been in the  stage of red
supergiants~[\cite{Takeda}]. Key features of
$\rho$\,Cas, which is often seen as a prototype of the YHG group,
are a long-known variability of the emission-absorption H$\alpha$
profile~[\cite{Sargent}], a high   matter loss rate
(it can reach up to $10^{-4}~\mathcal{M}_{\sun}$\,yr$^{-1}$, and
in the 2000 episode the star for about~200~days was loosing the
matter at a rate of up to
$3\times10^{-2}~\mathcal{M}_{\sun}$\,yr$^{-1}$~[\cite{Lobel2003}]),
as well as supersonic turbulence. Besides, the optical spectrum of
$\rho$\,Cas revealed a rarely observed feature---the splitting of
the  Ba\,II, Sr\,II, Ti\,II lines and other strongest absorptions
with a low lower-excitation potential level (an example is
presented in the bottom panel of Fig.~\ref{RhoCas}). Later, these features of the
$\rho$\,Cas spectrum  were investigated in detail through the
spectral monitoring~[\cite{Lobel1998,Lobel2006,rhocas1}],
including spectroscopy during the 2013 and 2017
ejections~[\cite{rhocas2, rhocas3}].
Spectral monitoring of $\rho$\,Cas performed at the BTA with the
spectral resolution of $R\ge60\,000$ allowed to estimate the
degree of variability of effective temperature:  the stellar
temperature variation in the range of \mbox{$5777$--$6744$~K} was
found~[\cite{rhocas1}].

A long series of high-quality spectral observations of $\rho$\,Cas
allows  to study the behavior of spectral feature profiles over
time, as well as the state of the velocity field in the extended
atmosphere and the circumstellar environment~[\cite{rhocas1, rhocas2}].
It is concluded on the closeness of velocities measured from the
absorptions of iron group ions, observed in the blue part of the
spectrum and the ones measured from the absorption components of
lines with inverse P\,Cyg profiles. The position of the absorption
components of lines with  inverse P\,Cyg profiles, reflecting the
presence of lumps of matter, infalling to the star with a velocity
of about $20$~km\,s$^{-1}$,  remained stable for all the
observation dates. Positions of strong Si\,II (multiplet~2)
absorptions and absorption components of the H$\alpha$ and
H$\beta$ lines changed insignificantly during the observational
period, staying, respectively, near the values of $63.7$ and
$70.5$~km\,s$^{-1}$.

%%%

Owing to the broad wavelength range, the velocity field is studied from
a record high number of single (several hundred in each
spectrum) and split absorptions (from 12 in the visible range to 89
in the shortwave range).
Radial velocity from weak symmetric
metal absorptions varies from date to date with the amplitude of
about~\mbox{$\pm7$~km\,s$^{-1}$} relative to the systemic velocity
\mbox{$V_{\rm sys}=-47$~km\,s$^{-1}$,}
which is a consequence of
low-amplitude pulsations in the stellar atmosphere, a
common property of YHGs. At some points, there is a dependence
of radial velocity versus line intensity, indicating
the existence of a velocity gradient in  deep layers of the stellar atmosphere.
For several phases, a difference  of velocities measured from the absorptions of
neutral atoms and ions was also found (at
$3$--$4$~km\,s$^{-1}$). Therefore, we have for the first time discovered a
stratification of velocities in the atmosphere of $\rho$\,Cas.

%%%

The authors of~[\cite{rhocas2}] showed that the long-wave component of the 
split absorptions  of Ba\,II, Sr\,II, Ti\,II and other strong lines with a low  lower-level excitation
potential is distorted by the stationary located emission, which
shifts the position of the line to the long wavelength region.
Thus, the longwave components of split absorptions are common
photospheric absorptions, the region of their formation and the
corresponding radial velocity, taking into account a distortion by
the stationary emission at certain times of observations do not
differ from those in single absorptions. $\rho$\,Cas is located on
the HRD at the boundary of the Yellow Void~[\cite{Jager1998}] dividing 
hypergiants and LBVs in the quiescent phase. On the boundary of the Yellow Void, 
the amplitude of pulsations of YHGs apparently  increases rather
greatly, which leads to an increased atmospheric instability and a
outburst~[\cite{Jager1998,Aret2017}]. In this regard,  note that in the
paper~[\cite{rhocas2}] the amplitude of the found
radial velocity variability from symmetric $V_r({\rm sym})$
absorptions exceeds $10$~km\,s$^{-1}$, which is higher than the
$V_r({\rm sym})$ variability values in our previous study of
$\rho$\,Cas~[\cite{rhocas1}].

%%%

In the summer of 2013, an ejection of matter occurred in the
$\rho$\,Cas system, at which the star's brightness decreased by
$0\fm5$ according to AAVSO. This ejection occurred only 12~years
after the previous one in 2000--2001. Therefore, we observe an
increased frequency of ejections in $\rho$\,Cas, which may
indicate an approach of the star to crossing the border of the
Yellow Void. After the 2013 ejection of matter  the spectrum
revealed TiO bands~[\cite{rhocas3}] and significant
H$\alpha$ profile type variations (see more details in the
article~[\cite{rhocas2}]). As we can see on the top
panel of Fig.~\ref{RhoCas}, before the flash, in
the 2010 spectrum the H$\alpha$ profile had a typical shape with
an emission in the short-wavelength wing. In the 2014 spectrum,
the core of the line is bifurcated for the first time. Wherein the
shortwave component of the bifurcated core is offset relative to
the systemic velocity by about~$-50$~km\,s$^{-1}$, which indicates
a rapid outflow of matter in the upper atmosphere after the
brightness minimum.

%%%
According to the set of radial velocity measurements in the spectra of $\rho$\,Cas
it is concluded that there is no correlation in the evolution of
H$\alpha$ profiles and split absorptions~[\cite{rhocas2}]. Wherein
a large velocity difference corresponding to the position of H$\alpha$ and an intermediate 
position of symmetric absorptions in the spectrum after the 2013 emission  (see~radial velocity 
measurement results in Fig.~5 of the article). Pulsation-type variability
with an amplitude of about~$10$~km\,s$^{-1}$ is inherent only to the symmetrical
low-to-moderate-intensity absorptions,  formation of which occurs in the deep atmospheric 
layers of the star.

%%%

A new dynamically unstable state of the gas envelope of the star
is registered in 2017, which manifested itself in a substantial
and quick variation of H$\alpha$ profile. In the spectrum of
February 13, 2017, the  H$\alpha$ profile before  the 2013
outburst got restored and a short-wave emission formed again.
However, a few months later, on August 6, 2017 and October 5,
2017, i.e. during an already new episode of matter ejection in
October 2017, we observe significant changes: the H$\alpha$ core
has split, its long-wave component has shifted into the long
wavelength region by about~$20$~km\,s$^{-1}$ relative to the
systemic velocity, indicating an infall of matter.

%%%

A significant variability is also recorded in the profiles of
strong split absorptions with a low  lower-level excitation
potential represented in particular on the bottom panel of
Fig.~\ref{RhoCas}. At the beginning of 2013, the
Ba\,II\,6141~\AA{} profile differed from the 2009 profile (i.e.
before the outburst) only in the variations in the shortwave wing,
where a weak emission of 2009 changed to an elongated absorption.
In the 2014 spectrum    (after a brightness minimum) we see a
significant shift and broadening of the short-wave envelope
component of the profile, indicating a velocity increase and an
increase in the gradient of the matter outflow velocity. A new
phenomenon in the spectrum was recorded after the 2017 outburst:
in February, the splitting of profiles of strong low excitation
absorptions into three components was registered for the first
time, which points to a change in the structure of the upper
atmosphere and the envelope of the star. We see the same picture
in the October 2017 spectra near the brightness minimum.

%%%

In the long-wavelength range   of the 2013 spectrum, weak envelope
emissions of iron group atoms are identified, while during the
2017 episode their intensity decreased until disappearance. In the
absence of emissions in the H and K nuclei of~Ca\,II lines,
envelope emissions of metals are constantly present in their
wings~[\cite{rhocas2}].

%%%

A three-component splitting of strong absorptions allows to talk
about a stratification in the uppermost layers of the outflowing
atmosphere and envelope of $\rho$\,Cas, i.e. about the occurrence
of two sub-envelopes with different kinematics. This new
phenomenon indicates the need to continue the spectral monitoring
of the star. When the signs of a new throw-off episode do appear,
observations have  to be carried  out 1--2 times a month, which
will allow to restore the temporal behavior of the kinematic state
of the atmosphere and envelope and evaluate the characteristic
time of variations.

\subsubsection{V509\,Cas, a hypergiant near the Yellow Void boundary}

\begin{figure}[ht!]
\includegraphics[angle=0,width=0.45\columnwidth, bb=20 80 480 680,clip]{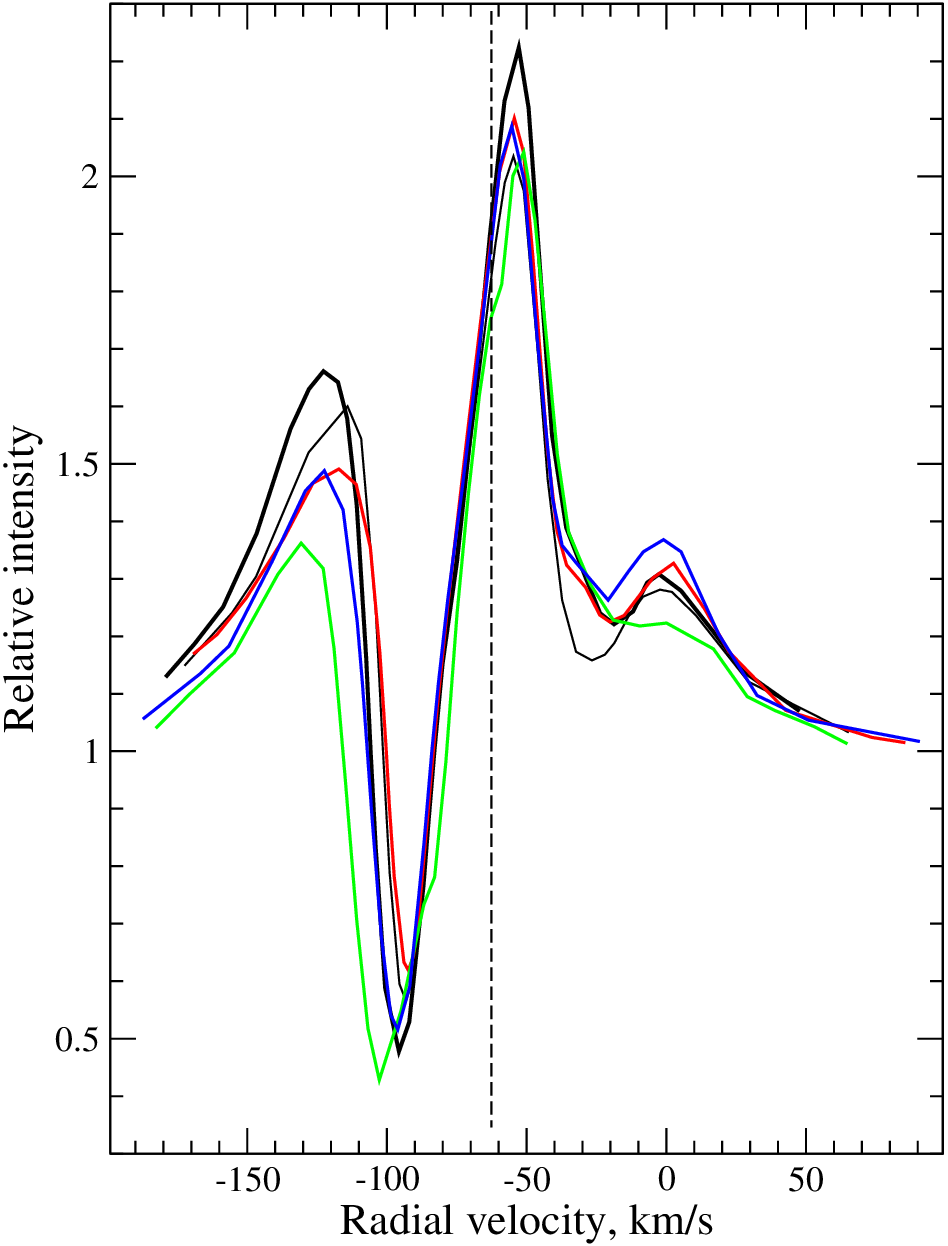}  %V509Cas_Halpha.eps
\caption{H$\alpha$ profile variability in the spectrum of
V509\,Cas: May 2, 1996---the green (dotted) line, October 1,
2014---the thick solid line, October 26, 2015---the blue line,
2017---the thin solid  line,
        April 6, 2018---the red line. The vertical dashed line indicates the systemic velocity value
        $V_{\rm sys}=-63$~km\,s$^{-1}$, accepted in~[\cite{V509Cas}].}
\label{V509CasHalpha}
\end{figure}

V509\,Cas is commonly considered to be a spectroscopic twin of the
hypergiant $\rho$\,Cas and therefore many authors compare their
spectra~[\cite{Lambert1981,Israelian1999,V509Cas}].
However, a detailed study of the spectra of these two hypergiants
with close fundamental parameters (mass, luminosity, stage of
evolution) revealed significant differences in both the spectra
and the kinematic state of the atmospheres, which indicates the
difference in physical processes causing an instability of the
atmospheres and envelopes. Spectral differences are manifested
primarily in the  difference of the H$\alpha$ profile type and its
variability. H$\alpha$ profile variations in the spectrum of
$\rho$\,Cas are mainly associated with the ejections. H$\alpha$
profile variations in Fig.~\ref{RhoCas} after the
2013 and 2017 ejections are particularly representative. In the
spectrum of V509\,Cas, the H$\alpha$ profile behaves more
smoothly, as we can see in
Fig.~\ref{V509CasHalpha}, the profile type,
position of its long-wave components are practically unchanged.
The H$\alpha$ absorption components  in this figure are located on
both sides of the  $V_{\rm sys}$ line, which indicates the
presence in the atmosphere and envelope of V509\,Cas of both the
outflowing and accreting matter. The shortwave component revealed
significant variations in the intensity, shift and in its width,
which is due to the variability of wind and instability of
conditions in the envelope, expanding with the velocity
of~\mbox{$33$--$40$~km\,s$^{-1}$.}

The authors of~[\cite{Israelian1999}], comparing the
spectra of V509\,Cas and $\rho$\,Cas in the near UV range, also
emphasized that these two stars are not spectral twins. Based on
the spectral features and evolutionary status, V509\,Cas is rather
closer to the hypergiant V1302\,Aql. First of all, H$\alpha$
profile types are close in their spectra, which indicates a
similarity of the circumstellar gas structure where these two-peak
emissions are formed. However,  these profiles still differ in the
details, as can be seen from a comparison of
Figs.~\ref{IRCHalpha}
and~\ref{V509CasHalpha}. H$\alpha$ profile
emissions in V1302\,Aql are stationary located and have a small
variable peak ratio~[\cite{IRC3}]. Emission peaks of
the H$\alpha$ profile in the spectra of V509\,Cas are many times
lower than in the case of V1302\,Aql. Variability of the long-wave
peak and the profile is in general insignificant. Short-wave
emission peak intensity varies, but it is constantly below the
long-wave one. No significant variations in the shape of metal
line profiles and the positions of their main components are
either found~[\cite{V509Cas}].

%%%

The optical spectrum of V509\,Cas is rich in emission features,
distorting the absorption profiles  not only of hydrogen but also metals, which is
clearly visible in the fragments of the spectrum of this star, presented in
Figs.~\ref{5150} and \ref{6200}. Figure~\ref{prof_7291_6300}
compares the profiles of the forbidden  [Ca\,II]\,7291 and
[O\,I]\,6300~\AA{}  emissions   in the spectra of V1302\,Aql and V509\,Cas.
The profiles of [Ca\,II]\,7291~\AA{} emission differ significantly in their
intensity, but they are single-peak in the spectra of both stars and have
close half-widths:  $\Delta V_r\approx40$~km\,s$^{-1}$.
Considering the presence of forbidden [Ca\,II] and [O\,I] emissions  in the spectrum
(Fig.~\ref{prof_7291_6300}), we confirm a conclusion of Aret et al.~[\cite{Aret510}] 
on the presence of a disk in the V509\,Cas system.
At the same time,  note that our one-peak profiles in the spectra of
V509\,Cas obtained with high spectral resolution
differ from the two-peak profiles in~[\cite{Aret510}].
It is also important that the width of these emissions is about 1.5--2 times lower than this
parameter in the spectrum of 3\,Pup, a star with the  B[e] phenomenon.
Besides, as seen in Fig.~5 from~[\cite{3Pup}], in the spectrum of 3\,Pup
the forbidden [Ca\,II] and [O\,I] emissions  have two-peak profiles,
which indicates the formation of these lines in the circumstellar
rotating Kepler disk~[\cite{Aret2016,Aret2017}].

%%%

The spectrum of V509\,Cas contains a feature, rarely observed in
the spectra of cool supergiants---highly excited forbidden
emissions of [N\,II]~5755, 6548 and 6584~\AA. Their presence in
the spectrum of such a cool single star is hard to explain. Having
forbidden [N\,II] lines in the spectrum of a yellow hypergiant
HR\,5171 has a natural explanation, the presence of a hot low-mass
companion in a close binary system with a common
envelope~[\cite{Chesneau2014}]. However, in case of V509\,Cas the presence 
of such a hot companion is reliably refuted in~[\cite{V509Cas}].

%%%

Back in 1965, Sargent~[\cite{Sargent65}] emphasized
that the presence of forbidden  [N\,II] emissions and complex
emission-absorption H$\alpha$ and H $\beta$ profiles indicates a
probable existence of a  hot envelope around V509\,Cas. In the
study~[\cite{Luck1975}] dedicated to an examination of chemical 
composition of the atmosphere of V509\,Cas, Luck adheres to the same position. 
Later, Lambert and Luck~[\cite{Lambert1978}] proposed several versions
of the excitation mechanism of the forbidden [N\,II] line in the
spectrum of V509\,Cas: dissipation of mechanical energy,
ionization due to UV radiation of hot stars populating an  H\,II
region in the  volume of the  Cep\,OB1 association, etc.

%%%

The authors of~[\cite{Genderen2015}] also believe
that the photons, necessary for the formation of highly excited
forbidden lines, can be produced by the yellow hypergiants
themselves, which demonstrate flashes on the time scale of 1--2
years with an amplitude of up to several magnitudes. An explosive
flash occurring due to a release during the recombination of the
hydrogen ionization energy, accumulated in the circumstellar
envelope, can provide the necessary high-energy photons. A good
example is a comprehensive study of the  hypergiant V509\,Cas
parameter variations~[\cite{Nieuwen2012}], which survived this kind 
of an episode around 1973, during which it threw out a great amount of mass, 
had a large diameter and a low temperature.

%%%

Analysis of the fundamental parameters of V509\,Cas over a large
time interval~[\cite{Nieuwen2012}] indicates that a hypergiant
like V1302\,Aql described a complex trajectory on the HRD.
According to~[\cite{Lambert1978}], at different stages of  evolution
the optical spectrum of the hypergiant varied in the spectral class range from
G to K.  Notwithstanding that, the luminosity class Ia was preserved. Currently,
the object has  approached the low-temperature boundary of the
Yellow Void~[\cite{Nieuwen2012}], which stimulates
the further spectral monitoring of this star.

\subsubsection{V1427\,Aql, a cooler analogue of V1302\,Aql}

The complexity of YHGs research is well-illustrated by the history
of  study of a cool supergiant V1427\,Aql\,=\,HD\,179821,
identified with an IR source IRAS\,19114+0002.
De~Jager~[\cite{Jager1998}], reckoning V1427\,Aql to the family of
yellow hypergiants, called this object a `small copy of
IRC+10420'. A combination of parameters of this star for a long
time did not allow to unambiguously specify its evolutionary
status (see~[\cite{ReddyHrivnak}]). However, quite convincing
evidence of V1427\,Aql belonging to the  YHG family has been
obtained in recent years~[\cite{Oudm2009,Sahin}]. Note that the
results in the paper~[\cite{Sahin}] are based on spectral
observations with an echelle spectrometer installed at the coude
focus of the McDonald Observatory 2.7-m telescope, and with the
6-m telescope in combination with the NES spectrograph. Oudmaijer
et al.~[\cite{Oudm2009}], analyzing the set of observed properties
of V1427\,Aql and V1302\,Aql came to a conclusion that  these two
stars are obvious yellow hypergiants.

%%%

Comparing the parameters of yellow hypergiants given in
Tables~\ref{objects} and~\ref{parameters}, we see that V1427\,Aql is
inferior to the members of the group only in terms of the matter
outflow rate. Its main parameters, including the intensity of the
O\,I\,7773~\AA{} triplet (and, therefore, luminosity too) and the
stellar wind velocity are close to those of V1302\,Aql. Modeling a
map of rotational transition profiles of the $^{12}$CO band, the
authors of the aforementioned paper~[\cite{Castro}] 
obtained  $V_{\rm exp}$ of about~$35$~km\,s$^{-1}$. These authors,
who noted that about $10^3$~years ago, the wind velocity of
V1427\,Aql reached an extremely high value of
$3\times10^{-3}\mathcal{M}_{\sun}$\,yr$^{-1}$,   consider this
object to be the closest relative of V1302\,Aql. According to
their calculations, the total mass of the envelope of V1427\,Aql
is four times the mass of the envelope of V1302\,Aql.

%%%

Figures~\ref{5150}, \ref{6200}, \ref{7250} and \ref{7450},
illustrate the fact that the spectrum of V1427\,Aql, as well as the spectrum of
$\rho$\,Cas predominantly contains  absorption features.
An exception are the emissions in the H$\alpha$ wings,
registered at individual moments, as well as a [Ca\,II]\,7291~\AA{} emission,
which, just like in the spectra of V1302\,Aql and
V509\,Cas, is single-peak---a narrow absorption overlapping
this emission is a telluric feature at the wavelength of
$\lambda=7291.4$~\AA.

\begin{figure}[ht!]
\includegraphics[angle=0,width=0.6\columnwidth, bb=15 75 545 675,clip]{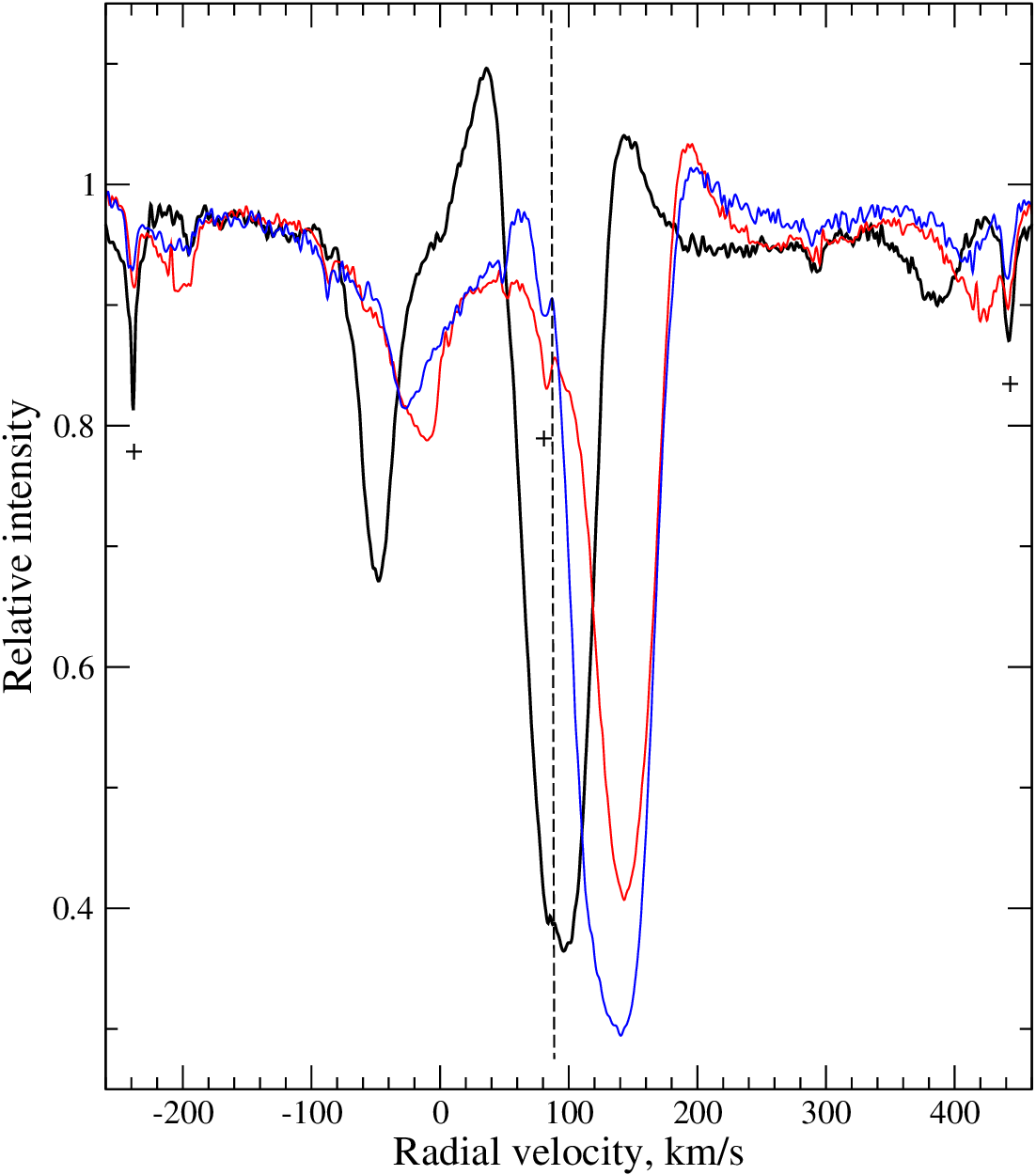}  %V1427Aql_Halpha.eps
\caption{Variability of the  H$\alpha$ profile in the spectrum of  V1427\,Aql:
September 24, 2010~---the blue  line,
May 30, 2013---the black thick line, October 9, 2013---a red line. The vertical dashed line
describes  the systemic velocity  $V_{\rm sys}=86$~km\,s$^{-1}$~[\cite{Likkel}].
Telluric features are marked by crosses.}
\label{V1427AqlHalpha}
\end{figure}

Figure~\ref{V1427AqlHalpha}  illustrates H$\alpha$
profile variability in the spectrum of V1427\,Aql. More data on
the variability of this line is available in Fig.~2 of Sahin et
al.~[\cite{Sahin}]. Emission in the profile wings, if
it appears, does not exceed 20\% of the local continuum. The type
of H$\alpha$ profile and its behavior in the spectrum of
V1427\,Aql is closer to that of $\rho$\,Cas. If the optical
spectrum of V1427\,Aql is close to the spectrum of $\rho$\,Cas,
then the extended circumstellar envelope of V1427\,Aql, as well as
the presence of a weak [Ca\,II] emission,  getting  formed in an
optically thin envelope, makes this star more akin to  V1302\,Aql.

%%%
The authors of~[\cite{Sahin}] obtained critical
arguments supporting the status of a massive far-evolved star for
V1427\,Aql. Namely, a large   nitrogen and sodium excess, a high
microturbulent velocity value and supersonic macroturbulent
velocity. As it can be seen from
Table~\ref{parameters}, high luminosity of the star
corresponds to a large equivalent width of the infrared oxygen
triplet: \mbox{$W_{\lambda}(7773)=2.7$~\AA}. Note that such a high
$W_{\lambda}(7773)$ value   is measured in the spectrum of the
star, for which a decreased oxygen abundance in the atmosphere was
determined based on weak absorptions~[\cite{Sahin}].
In general, reinforcing the conclusion of de~Jager~[\cite{Jager1998}], we can say that
V1427\,Aql is a cooler and less studied analogue of the hypergiant
V1302\,Aql (IRC+10420), while the latter is considered a
cornerstone in the study of late phases of evolution of massive
stars~[\cite{Clark2014}].

%%%

It is obvious that our Galaxy contains also weak in the optical
range, and therefore so far little studied YHGs. For example,
according to the authors of~[\cite{Clark2014}], a
distant source IRAS\,18357$-$0604 is another twin of IRC$+$10420.
Having high luminosity, the central star of
\mbox{IRAS\,18357$-$0604} is not reachable by the means of high
resolution spectroscopy due to the remoteness of the object and a
large extinction in the circumstellar medium. However, according
to~[\cite{Clark2014}], its long-wavelength spectrum,
as well as the spectrum of V1302\,Aql is saturated with asymmetric
two-peak Paschen series H\,I line emissions, Ca\,II,
low-excitation metal lines. The presence of nitrogen N\,I\,8630
and 8729~\AA{} emissions in the spectrum of IRAS\,18357$-$0604 is
an evidence of nitrogen enrichment of the atmosphere and the
envelope, as in the case of IRC+10420. A particularly interesting
object from the YHGs family is known in the southern sky, this is
a hypergiant V766\,Cen\,$=$\,HR\,5171 in a binary interacting
system with a common envelope~[\cite{Chesneau2014}].

\subsection{Possible relationship of YHGs and supergiants with B[e] phenomenon}

The above YHGs are studied by various methods in all wavelength
ranges. However, their evolutionary status is still not quite
certain, since they can be attributed either to the stars, that
evolve to RSG, or to stars at the  stage after the RSG. Upon
completion of the burning of hydrogen in the core, the descendants
of RSGs make their way to the hot area of HRD. Typical
representatives of RSGs are the cool red supergiants VY\,CMa,
Betelgeuse and $\mu$\,Cep with the spectral classes
Sp\,=\,M2--M5\,Ia. A key moment that can help solving the problem
of evolutionary status is the presence or absence of a
circumstellar envelope and the features of its structure, since
the envelope, reflecting the history of matter loss variation
parameters is a kind of a recording of the history of the star. A
long-standing mystery in the YHG family is a lack of a dust
envelope in $\rho$\,Cas with a matter loss rate of
$\ge10^{-4}\mathcal{M}_{\sun}$\,yr$^{-1}$ (and
over~$10^{-3}\mathcal{M}_{\sun}$\,yr$^{-1}$ during the ejection
episodes~[\cite{Lobel2003}]). An IR excess which is
natural to expect of a star of such a high luminosity with a high
matter loss rate, has for a long time  not been detected in
$\rho$\,Cas (see the references in~[\cite{Shuster2006,Shenoy2016}]).
Wherein the presence of a circumstellar gas envelope appears in
the CO molecule bands (see the papers~[\cite{Lambert1981,Lobel2006}] 
and references therein), as well as in the components of the
profiles of Na\,I, K\,I  and low-excited lines of a number of
ions~[\cite{rhocas2}]. The most natural factor
explaining the presence (or absence) of a powerful dust envelope
and its properties may be a difference in the luminosity of the
star, in  the matter loss rate associated with it, and ultimately,
in the initial mass and the features of passage of an evolutionary
phase. Apparently, as suggested by the authors of~[\cite{Quintana2008}], 
we observe $\rho$\,Cas at an earlier stage of evolution than that of hypergiants with a
large  IR-flux excess.

%%%

In connection with the problem of fixing the evolutionary stage in
individual YHGs, an urgent task is to study the evolutionary
relationship between them and supergiants with  B[e] phenomenon.
The possibility of such a connection already follows from the
relative position of these two groups of star types on the HRD.
Aret et al.~[\cite{Aret2017}] analyzed this kind of
connection between two groups of massive HLSs, based on a
combination of  IR flux excesses with the presence in the spectra
of both groups of stars of forbidden  [O\,I] and [Ca\,II]
emissions, the formation of which occurs in an optically dense
circumstellar disk in supergiants with a B[e] phenomenon and the
subgroup of hot YHGs. Note that the [Ca\,II]\,7291~\AA{} emission
is clearly visible  in the October 1, 2014 spectrum of $\rho$\,Cas
in Fig.~\ref{prof_7291_6300}. However, this feature
did not exceed the continuum level in our spectra before the
outburst and after 2017, when the star was in the quiescent state
and its flux in the local continuum significantly exceeded the
flux in the [Ca\,II]\,7291~\AA{} emission. We  registered no
emission in the [O\,I]\,6300~\AA{} line in the spectra of
$\rho$\,Cas  in any of our observational sets.

%%%

The optical spectrum of V509\,Cas is similar to the spectrum of
the star 3\,Pup with the B[e] phenomenon. In the spectrum of this
coolest representative (with the spectral class Sp\,$=$\,A4\,Iabe)
of the family of supergiants with the B[e] phenomenon, a
two-peaked H$\alpha$  with a stronger red component, forbidden
[O\,I]\,1F\,6300, 6364~\AA\ lines emissions,  a
[Ca\,II]\,1F\,7291, 7324~\AA\ doublet and emissions in some Fe\,II
lines are registered. According to~[\cite{3Pup}], in
the 3\,Pup spectrum only the Mg\,II\,4481~\AA{} line can be
considered photospheric, while in the Fe\,II lines a envelope
contribution is obvious, it gives profiles a specific shape: wings
are raised by the emissions. The same feature is inherent to the
profiles  in the spectrum of V509\,Cas too, in which the wings of
even the weak lines are distorted by the emissions, which is
clearly visible on the top panel of Fig.~\ref{7450}
at the Fe\,II\,7462~\AA\ line with the depth not exceeding~0.2 of
the local continuum level.

%%%

Note that the authors of~[\cite{Chesneau2014}] suggested that
the aforementioned hypergiant V766\,Cen   can evolve into a
3\,Pup-type system, as it moves into the hot region of the
HRD. Obviously, the hypothesis of a possible evolutionary
relationship between the two groups of HLSs requires additional research, 
including that in the framework of the stellar-astrometric approach.

\section{Conclusions}\label{conclus}

We have presented the results of a long term spectral monitoring
of yellow hypergiants of the northern celestial hemisphere.
Fundamental parameters of stars were determined based on the
homogeneous spectroscopy data with high spectral resolution. An IR
oxygen triplet OI\,7773~\AA{} with extreme  equivalent width
values served as a luminosity criterion: the considered stars have
an average value of $\log L/L_{\sun}=5.43\pm0.14$.

%%%

Based on  the ensemble of  detailed positional measurement data we
estimated the circumstellar envelope expansion rate in the range
of 11$\div$40\,km\,s$^{-1}$. Based on the weak absorptions
the amplitude of pulsations in a narrow range of values
\mbox{$\Delta V_r$=7$\div$11\,km\,s$^{-1}$} was determined for four
objects.

%%%

We demonstrated a diversity of observed spectral features in stars
of extremely high luminosity, compactly located at the top of the
Hertzsprung--Russell diagram, namely, the presence (or absence) of
permitted and forbidden emissions, emission components of complex
profiles, temporal features of behavior of spectral features of
various nature.

%%%

High spectral resolution monitoring efficiency is shown for the
detection of dynamic state variability at different depths of the
extended atmosphere and circumstellar envelope of the hypergiants.
In particular, due to many years of monitoring of the hypergiant
V1302\,Aql we conclude that it is approaching the low temperature
boundary of the Yellow Void, which indicates the relevance of
continuing to monitor this star.

%%%

The reliability of the hypergiant status for V1427\,Aql and the
lack of a companion in the V509\,Cas was established. Monitoring
of $\rho$\,Cas allowed to fix the dynamic instability of the upper
atmosphere of the star and for the first time to detect a
stratification of its gas envelope during the 2017 matter ejection
episode. Due to the more frequent ejection episodes, we emphasize
absolute importance of continuing the spectral monitoring of
$\rho$\,Cas.

\begin{acknowledgements}
The author is grateful to his co-authors who participated in the implementation of the
supergiant spectroscopy program at the BTA and in the preparation of
joint publications. 
The author thanks   The Russian Foundation for Basic
Research for the financial support (project 18--02--00029\,a).
We made  use of the astronomical databases
SIMBAD, SAO/NASA ADS, AAVSO and VALD.
\end{acknowledgements}


\newpage
\begin{thebibliography}{}

\bibitem{Hump1979} 1. R.~M. {Humphreys} and K.~{Davidson}, \apj\  \textbf{232}, 409 (1979).

\bibitem{Jager1998} 2.  C.~{de Jager}, \araa\ \textbf{8}, 145 (1998).

\bibitem{Hump2017} 3. R.~M. {Humphreys}, K.~{Davidson}, D.~{Hahn}, et~al., \apj\
\textbf{844}, 40 (2017).

\bibitem{Aret510} 4. A.~{Aret}, M.~{Kraus}, I.~{Kolka}, and G.~{Maravelias},
  ASP Conf. Series {\bf 510}, 162 (2017).

\bibitem{Gaia} 5. {Gaia Collaboration}, A.~G.~A. {Brown}, A.~{Vallenari}, et~al.,
\aaa\ \textbf{616}, A1 (2018).


\bibitem{Castro} 6. A.~{Castro-Carrizo}, G.~{Quintana-Lacaci}, V.~{Bujarrabal},
et~al., \aaa\ \textbf{465}, 457 (2007).

\bibitem{Jones93} 7. T.~J. {Jones}, R.~M. {Humphreys}, R.~D. {Gehrz}, et~al., \apj\
\textbf{411}, 323 (1993).

\bibitem{Shuster2006} 8. M.~T. {Schuster}, R.~M. {Humphreys}, and M.~{Marengo}, \aj\
\textbf{131}, 603 (2006).

\bibitem{Meynet2007} 9. G.~{Meynet}, P.~{Eggenberger}, and A.~{Maeder}, IAU Symposium {\bf 241}, 13 (2007).

\bibitem{Jura2001} 10. M.~{Jura}, T.~{Velusamy}, and M.~W. {Werner}, \apj\ \textbf{556}, 408 (2001).

\bibitem{Massey2013} 11.  P.~{Massey}, New Astron. Rewies \textbf{57}, 14 (2013).

\bibitem{Nieuwen2012} 12. H.~{Nieuwenhuijzen}, C.~{De Jager}, I.~{Kolka}, et~al., \aaa\
\textbf{546}, A105 (2012).


\bibitem{Jager2001} 13.  C.~{de Jager}, A.~{Lobel}, H.~{Nieuwenhuijzen}, and R.~{Stothers},
\mnras\ \textbf{327}, 452 (2001).

\bibitem{Boyarchuk} 14. A.~A. {Boyarchuk}, I.~{Gubeny}, I.~{Kubat}, et~al., Astrophysics
\textbf{28}, 202 (1988).

  %15
\bibitem{IRC1} 15. V.~G. {Klochkova}, E.~L. {Chentsov}, and V.~E. {Panchuk}, \mnras\
\textbf{292}, 19 (1997).

\bibitem{Sahin} 16.  T.~{{\c{S}}ahin}, D.~L. {Lambert}, V.~G. {Klochkova}, and
V.~E.~{Panchuk}, \mnras\ \textbf{461}, 4071 (2016).


\bibitem{nes1} 17.  V.~E. {Panchuk}, V.~G. {Klochkova}, M.~V. {Yushkin}, and I.~D.
{Najdenov}, Journal of Optical Technology \textbf{76}, 87 (2009).

\bibitem{nes2}  18. V.~E. {Panchuk}, V.~G. {Klochkova}, and M.~V. {Yushkin}, \arep\
\textbf{61}, 820 (2017).

\bibitem{echelle} 19.  M.~V. {Yushkin} and V.~G. {Klochkova}, Preprint No.~206, (Special
Astrophysical Observatory of RAS, 2004).


\bibitem{dech} 20. G.~A. {Galazutdinov}, Preprint No.~92, (Special Astrophysical
Observatory of RAS, 1992).

\bibitem{atlas} 21.  E.~L. {Chentsov}, V.~G. {Klochkova}, and N.~S.~{Tavolganskaya},
\bsao\ \textbf{48}, 25 (1999).


\bibitem{atlasFG} 22.  V.~G. {Klochkova}, G.~{Zhao}, V.~E. {Panchuk}, and S.~V.
{Ermakov}, Chin. J. Astron. and Astrophys. \textbf{4}, 279 (2004).

\bibitem{Kastner} 23.  J.~H. {Kastner} and D.~A. {Weintraub}, \apj\ \textbf{452}, 833
(1995).

\bibitem{Xu2019} 24.  S.~{Xu}, B.~{Zhang}, M.~J. {Reid}, et~al., \apj\ \textbf{875}, 114
(2019).

\bibitem{V1648Aql} 25. V.~G. {Klochkova} and N.~S. {Tavolzhanskaya}, \ab\  \textbf{74}, 277 (2019).


\bibitem{rhocas2} 26.  V.~G. {Klochkova}, V.~E. {Panchuk}, and N.~S.~{Tavolzhanskaya}, \arep\ \textbf{62}, 623 (2018).
  \textbf{95}, 659 (2018).

\bibitem{V509Cas} 27. V.~G. {Klochkova}, E.~L. {Chentsov}, and V.~E. {Panchuk}, \ab\ \textbf{74}, 41 (2019).

\bibitem{IRC3} 28. V.~G. {Klochkova}, E.~L. {Chentsov}, A.~S. {Miroshnichenko}, et~al., \mnras\ \textbf{459}, 4183 (2016).

\bibitem{Luck2014} 29.  R.~E. {Luck}, \aj\ \textbf{147}, 137 (2014).

\bibitem{Takeda1998} 30. Y.~{Takeda} and M.~{Takada-Hidai}, \pasj\ \textbf{50}, 629 (1998).

\bibitem{Kovtyukh} 31.  V.~V. {Kovtyukh}, N.~I. {Gorlova}, and S.~I. {Belik}, \mnras\
\textbf{423}, 3268 (2012).


\bibitem{Lobel1992} 32.  A.~{Lobel}, L.~{Achmad}, C.~{de Jager}, and H.~{Nieuwenhuijzen},
\aaa\ \textbf{256}, 159 (1992).

\bibitem{Hump1997} 33.  R.~M. {Humphreys}, N.~{Smith}, K.~{Davidson}, et~al., \aj\
\textbf{114}, 2778 (1997).

\bibitem{IRC2} 34.  V.~G. {Klochkova}, M.~V. {Yushkin}, E.~L. {Chentsov}, and V.~E. {Panchuk},   \arep\ \textbf{46}, 139 (2002).
  \textbf{79}, 158  (2002).

\bibitem{Oudm2009} 35.  R.~D. {Oudmaijer}, B.~{Davies}, W.~J. {de Wit}, and M.~{Patel},
  ASP Conf. Series {\bf 412}, 17 (2009).

\bibitem{Bakker97}  36. E.~J. {Bakker}, E.~F. {van Dishoeck}, L.~B.~F.~M. {Waters}, and
  T.~{Schoenmaker}, \aaa\ \textbf{323}, 469 (1997).

\bibitem{Loup} 37.  C.~{Loup}, T.~{Forveille}, A.~{Omont}, and J.~F. {Paul}, \aas\
\textbf{99}, 291  (1993).

\bibitem{Egg1} 38.  V.~G. {Klochkova}, R.~{Szczerba}, and V.~E. {Panchuk}, \alet\ \textbf{26}, 439 (2000).
  
\bibitem{Aret2017} 39.  A.~{Aret}, M.~{Kraus}, I.~{Kolka}, and G.~{Maravelias},
  ASP Conf. Series {\bf 508}, 357 (2017).

\bibitem{rhocas3}  40. M.~{Kraus}, I.~{Kolka}, A.~{Aret}, et~al., \mnras\ \textbf{483},
3792 (2019).

\bibitem{Giguere} 41.  P.~T. {Giguere}, N.~J. {Woolf}, and J.~C. {Webber}, \apj\
   \textbf{207}, L195 (1976).

\bibitem{Mutel}  42.  R.~L. {Mutel}, J.~M. {Benson}, J.~D. {Fix}, and J.~C.~{Webber},
\apj\ \textbf{228}, 771 (1979).

\bibitem{Oudm1998} 43.  R.~D. {Oudmaijer}, \aas\ \textbf{129}, 541 (1998).

\bibitem{Oudm1996} 44.  R.~D. {Oudmaijer}, M.~A.~T. {Groenewegen}, H.~E.~{Matthews},
et~al., \mnras\ \textbf{280}, 1062 (1996).


\bibitem{Hump2002}  45. R.~M. {Humphreys}, K.~{Davidson}, and N.~{Smith}, \aj\
\textbf{124}, 1026 (2002).

\bibitem{Quintana13} 46.  G.~{Quintana-Lacaci}, M.~{Ag{\'u}ndez}, J.~{Cernicharo}, et~al.,
\aaa\ \textbf{560}, L2 (2013).

\bibitem{Takeda}  47. Y.~{Takeda} and M.~{Takada-Hidai}, \pasj\ \textbf{46}, 395 (1994).

\bibitem{Sargent} 48.  W.~L.~W. {Sargent}, \apj\ \textbf{134}, 142 (1961).

\bibitem{Lobel2003} 49. A.~{Lobel}, A.~K. {Dupree}, R.~P. {Stefanik}, et~al., \apj\
\textbf{583}, 923 (2003).

\bibitem{Lobel1998}  50. A.~{Lobel}, G.~{Israelian}, C.~{de Jager}, et~al., \aaa\
\textbf{330}, 659 (1998).

\bibitem{Lobel2006} 51.  N.~{Gorlova}, A.~{Lobel}, A.~J. {Burgasser}, et~al., \apj\
\textbf{651}, 1130  (2006).

\bibitem{rhocas1} 52.  V.~G. {Klochkova}, V.~E. {Panchuk}, N.~S.~{Tavolzhanskaya}, and I.~A.~{Usenko},  \arep\ \textbf{58}, 101 (2014).

\bibitem{Lambert1981} 53.  D.~L. {Lambert}, K.~H. {Hinkle}, and D.~N.~B. {Hall}, \apj\
\textbf{248}, 638 (1981).

\bibitem{Israelian1999} 54.  G.~{Israelian}, A.~{Lobel}, and M.~R. {Schmidt}, \apj\
\textbf{523}, L145  (1999).

\bibitem{3Pup} 55. E.~L. {Chentsov}, V.~G. {Klochkova}, and A.~S.~{Miroshnichenko}, \ab\ \textbf{65}, 150 (2010).
   159 (2010).

\bibitem{Aret2016} 56.  A.~{Aret}, M.~{Kraus}, and M.~{{\v{S}}lechta}, \mnras\ \textbf{456}, 1424 (2016).

\bibitem{Chesneau2014}  57.  O.~{Chesneau}, A.~{Meilland}, E.~{Chapellier}, et~al., \aaa\
\textbf{563}, A71  (2014).

\bibitem{Sargent65} 58. W.~L.~W. {Sargent}, The Observatory \textbf{85}, 33 (1965).

\bibitem{Luck1975} 59. R.~E. {Luck}, \apj\ \textbf{202}, 743 (1975).

\bibitem{Lambert1978} 60.  D.~L. {Lambert} and R.~E. {Luck}, \mnras\ \textbf{184}, 405 (1978).

\bibitem{Genderen2015} 61.  A.~M. {van Genderen}, H.~{Nieuwenhuijzen}, and A.~{Lobel}, \aaa\
\textbf{583}, A98 (2015).

\bibitem{ReddyHrivnak} 62.  B.~E. {Reddy} and B.~J. {Hrivnak}, \aj\ \textbf{117}, 1834 (1999).

\bibitem{Likkel}  63.  L.~{Likkel}, A.~{Omont}, M.~{Morris}, and T.~{Forveille}, \aaa\
\textbf{173},  L11 (1987).

\bibitem{Clark2014}  64.  J.~S. {Clark}, I.~{Negueruela}, and C.~{Gonz{\'a}lez-Fern{\'a}ndez}, \aaa\ \textbf{561}, A15 (2014).

\bibitem{Shenoy2016} 65.  D.~{Shenoy}, R.~M. {Humphreys}, T.~J. {Jones}, et~al., \aj\
\textbf{151}, 51 (2016).

\bibitem{Quintana2008} 66.  G.~{Quintana-Lacaci}, V.~{Bujarrabal}, and A.~{Castro-Carrizo},
\aaa\ \textbf{488}, 203 (2008).

\end{thebibliography}
\end{document}